\documentclass[%
 reprint,
superscriptaddress,
 amsmath,amssymb,
 aps,
floatfix,
]{revtex4-2}  % see https://web.mit.edu/8.13/www/revtex4-command-summary.pdf

\newcommand{\fnm}[1]{#1}
\newcommand{\sur}[1]{\surname{#1}}
\newcommand{\orgname}[1]{#1}
\newcommand{\city}[1]{#1}
\newcommand{\country}[1]{#1}

\usepackage[version=4]{mhchem}

\usepackage{float}
\usepackage[english]{cleveref}
\usepackage{csquotes}
\usepackage[inline]{enumitem}

\usepackage[dvipsnames]{xcolor}         % citation colors
\usepackage{booktabs}                   % for \midrule

\usepackage{derivative}
\usepackage{xfrac}

\usepackage[per-mode=symbol-or-fraction, uncertainty-mode=separate]{siunitx}
\DeclareSIUnit{\mK}{\milli\K}
\DeclareSIUnit{\yJ}{\yocto\J}

\begin{document}

\title{Josephson microwave photon detector operating at 0.7 K}

\author{\fnm{D.~A.} \sur{Ladeynov}}
\affiliation{\orgname{Nizhny Novgorod State Technical University n.~a. R.~E.~Alekseev}, \city{Nizhny Novgorod}, \country{Russia}}
\affiliation{\orgname{N.~I.~Lobachevsky State University of Nizhny Novgorod}, \city{Nizhny Novgorod}, \country{Russia}}
\affiliation{\orgname{Institute for Physics of Microstructures of RAS}, \city{Nizhny Novgorod}, \country{Russia}}
% \email[<optional text>]{author@any.edu}

\author{\fnm{A.~L.} \sur{Pankratov}}
\affiliation{\orgname{Nizhny Novgorod State Technical University n.~a. R.~E.~Alekseev}, \city{Nizhny Novgorod}, \country{Russia}}
\affiliation{\orgname{N.~I.~Lobachevsky State University of Nizhny Novgorod}, \city{Nizhny Novgorod}, \country{Russia}}
\affiliation{\orgname{Institute for Physics of Microstructures of RAS}, \city{Nizhny Novgorod}, \country{Russia}}
% \email[<optional text>]{author@any.edu}

\author{\fnm{L.~S.} \sur{Revin}}
\affiliation{\orgname{Nizhny Novgorod State Technical University n.~a. R.~E.~Alekseev}, \city{Nizhny Novgorod}, \country{Russia}}
\affiliation{\orgname{Institute for Physics of Microstructures of RAS}, \city{Nizhny Novgorod}, \country{Russia}}

\author{\fnm{A.~V.} \sur{Gordeeva}}
\affiliation{\orgname{Nizhny Novgorod State Technical University n.~a. R.~E.~Alekseev}, \city{Nizhny Novgorod}, \country{Russia}}
\affiliation{\orgname{Institute for Physics of Microstructures of RAS}, \city{Nizhny Novgorod}, \country{Russia}}

\author{\fnm{A.~V.} \sur{Chiginev}}
\affiliation{\orgname{Nizhny Novgorod State Technical University n.~a. R.~E.~Alekseev}, \city{Nizhny Novgorod}, \country{Russia}}
\affiliation{\orgname{N.~I.~Lobachevsky State University of Nizhny Novgorod}, \city{Nizhny Novgorod}, \country{Russia}}
\affiliation{\orgname{Institute for Physics of Microstructures of RAS}, \city{Nizhny Novgorod}, \country{Russia}}

\author{\fnm{S.~A.} \sur{Razov}}
\affiliation{\orgname{Nizhny Novgorod State Technical University n.~a. R.~E.~Alekseev}, \city{Nizhny Novgorod}, \country{Russia}}
\affiliation{\orgname{N.~I.~Lobachevsky State University of Nizhny Novgorod}, \city{Nizhny Novgorod}, \country{Russia}}

\author{\fnm{E.~V.} \sur{Il'ichev}}
\affiliation{\orgname{Leibniz Institute of Photonic Technology}, \city{Jena}, \country{Germany}}

\begin{abstract}
We predict that the threshold detectors based on \ce{Al} Josephson junctions, with critical currents below \qty{100}{\nA}, exhibiting a phase diffusion regime, can be exploited for the microwave photon detection both at 17 mK and 700 mK. We demonstrate a detection of two- and one-photon energies at \qty{5}{GHz} with 90 \% and 15 \% efficiency and dark count time of about 0.1 s and 0.01 s, respectively. The observed weak temperature dependence of the detector's performance in the sub-kelvin range fully confirms its phase-diffusion mode of operation. On the other hand, these results show that inevitable thermal fluctuations are not the main source of the detector noise. Consequently, there is still a room to optimize the detector's performance. These results are important for axion search experiments in the range of 5-25 GHz (20-100 $\mu$eV).
\end{abstract}


\maketitle
\section{Introduction}
Modern developments in quantum communications and quantum information processing devices benefit greatly from the efficient detection of individual photons.
However, widely used conventional single-photon detectors, such as superconducting nanowires \cite{EsmaeilZadeh2021May, Khasminskaya2016Nov, Kahl2015Jun} or transition-edge devices \cite{Irwin2005Jul, Fortsch2015Apr}, require a sufficiently high incident photon energy to modify the state of the detector.
For the microwave frequency range, the corresponding energy is rather low, about \qty{10}{\yJ}.
Consequently, practicable single photon detectors for microwave field have not been implemented yet.
A natural idea for realizing such a detector is to exploit quantum Josephson circuits.
Indeed, superconducting qubits, being two-level quantum systems with characteristic energy in the microwave scale, can be excited by absorbing an incident photon.
Various types of qubit- and quantum-dot-based photon detectors have been proposed and implemented \cite{Johnson2010Sep,Inomata2016Jul, Wong2017Jan, Besse2018Apr, Royer2018May, Kono2018Jun, Lescanne2020May, Albertinale2021, Grimsmo2021Mar, Dixit2021Apr, Khan2021Aug, Balembois2024Jan, Haldar2024Feb} with efficiencies approaching  $\approx 60\%$ \cite{Lescanne2020May, Balembois2024Jan}.

\begin{figure}[h!]
    \centering
    \includegraphics[width=\linewidth, keepaspectratio]{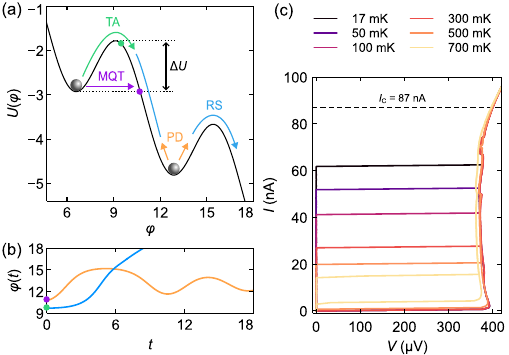}
    \caption[]{Phase particle regimes and current–voltage characteristics (IVCs).
    \begin{enumerate*}[label=(\alph*)]
        \item Regimes in a tilted potential.
              Here, TA means thermal activation, MQT - macroscopic quantum tunneling, RS - switching to the running state with finite voltage and PD - phase diffusion regime with re-trapping.
        \item Phase evolution.
              If a particle escapes due to TA, it gains larger potential energy (green dot), and thus has a higher probability of switching to RS.
              If a particle tunnels under the barrier (violet dot), it has smaller potential energy and higher probability of re-trapping, leading to quantum PD.
        \item IVC of SIS2; dashed line is theoretical critical current.
    \end{enumerate*}}
    \label{fig:potential}  % TODO: refer
\end{figure}

As a photon detector we use a current biased Josephson junction (CBJJ), which was also utilized before as the phase qubit \cite{Martinis2002Aug}.
The phase across the junction for such devices is associated with the dynamics of a quantum particle in a washboard potential \cite{Barone1982}, see Fig.~\ref{fig:potential}a.
In the initial state of the detector, the particle is trapped in one of the local potential minima and the voltage across the CBJJ is zero.
After the incident microwave field delocalizes the trapped particle, it escapes to the \enquote{running state} (RS), providing a finite voltage across the CBJJ.
This voltage is in the millivolt range and can be easily measured.

However, to realize maximum sensitivity, the CBJJ must be biased relatively close to the critical current value.
Due to fluctuations, such initialization causes uncontrolled transitions to a voltage state, corresponding to the dark count rate.
The trade-off between initialization point and resolution, which is common for threshold-type detectors, is clearly seen for this particular realization.
In fact, this is the main obstacle to implementing an efficient single photon detector for the microwave range. Importantly, the relatively long lifetime of the threshold detector in the initial (zero-voltage) state is a key issue for Dark Matter axion-type particle searches \cite{Lamoreaux2013Aug, Irastorza2018Sep, Sikivie2021Feb,PhysRevLett.124.101303, Crescini2020May, Kwon2021May, Adair2022Oct, Sushkov2023May, Graham2024Feb}. Moreover, it has recently been predicted \cite{Buschmann2022Feb} that one of the most interesting ranges for the search for axions is 20-100 $\mu$eV, corresponding to the 5-25 GHz range, covered by the detectors considered here.

Various approaches have been used to optimize the CBJJ threshold detector \cite{Chen2011Nov, Peropadre2011Dec, Addesso2012Jan, Poudel2012Nov, Kuzmin2018Jun, Guarcello2019Apr, Guarcello2019May, Yablokov2020Jul, PiedjouKomnang2021Jan, Yablokov2021Jul, Golubev2021Jul, Ladeynov2023Jun, Stanisavljevic2024Aug}.
Interestingly, the theoretically estimated lifetime in the zero voltage state for the studied threshold detector is sufficiently shorter than the experimental one. To explain this inconsistency, it has been proposed that CBJJ is in the so-called phase-diffusion regime \cite{Pankratov2022May}.

The main feature of the phase diffusion (PD) regime, in a handwaving manner, can be commented on as follows.
After the escape from the local minimum of the washboard potential via macroscopic quantum tunneling (MQT) or thermal activation (TA), the phase can be re-trapped into another minimum with a relatively high probability, see Fig.~\ref{fig:potential}a,b.
Consequently, in the PD regime the CBJJ much rarely switches to the voltage state due to re-trapping, which effectively increases its lifetime in the initial state.
However, in the case of arriving photon, switching into the running state without re-trapping may occur, due to the induced current pulse that significantly exceeds the detector threshold.
While the main characteristics of the phase diffusion regime are well described \cite{Martinis1989Oct, Kautz1990Dec, Koval2004Aug, Kivioja2005Jun,  Mannik2005Jun, Krasnov2005Oct, Krasnov2007Dec, Fenton2008Aug, Yoon2011May, Yu2011Aug, Longobardi2011Nov, Longobardi2012Aug, Yu2012Dec, Revin2020Jun}, its application for photon counting is unknown.
In this paper, after a detailed investigation of the switching dynamics of CBJJs, we demonstrate a single and double photon detection at 5 GHz with an efficiency of about 15 \% and 90 \%, respectively. Importantly, increasing the temperature by more than 40 times (from \qty{17}{\mK} to \qty{700}{\mK}, see below) does not lead to degradation of the detector performance for a properly chosen sample. In practice, this counter-intuitive result means that inevitable thermal fluctuations can be ”compensated” by the phase diffusion mechanism.

\section{Theoretical model}
The dynamics of the CBJJ can be represented as a phase particle of mass $C$ (capacitance) moving under the effect of damping 1/$R$ (with the normal state resistance $R$) in a tilted periodic potential $U = -E_\text{J} (i\phi + cos \phi)$, Fig.~\ref{fig:potential}a, according to the resistively and capacitively shunted junction (RCSJ) model. Here, $E_\text{J} = \hbar I_\text{C} / (2e)$ is the Josephson energy ($I_\text{C}$ is the CBJJ critical current, $\hbar$ is the reduced Planck constant, $e$ is the electron charge), $i = I/I_\text{C}$ is the normalized bias current. Previously, we described several scenarios (MQT, RS, PD) of phase behavior \cite{Pankratov2024May}. The transition to these scenarios is connected with a phase particle dynamics, escaping from a potential well. Due to the effect of noise, this state is metastable, and the particle can escape across the barrier due to TA, or tunnel through the barrier at lower temperatures (MQT). 

The lifetime (time of staying in a potential well) is described by the following set of approximate expressions (Kramers and MQT theories) $\tau_\text{TA}$ (TA regime) and $\tau_\text{Q}$ (MQT regime) \cite{Voss1981Jul, Martinis1988Aug, Oelsner2013Sep, Blackburn2016Feb}:
\begin{equation}
    \left[ 
        \begin{gathered}
            \tau_\text{TA} = \frac{2\pi}{\omega_0}a_\text{TA}^{-1} e^{\frac{\Delta{}U}{k_\text{B} T}}, \hfill T \geq T_\text{Q}
            \hfill
            \\
            \tau_\text{Q} = \frac{2\pi}{\omega_0} \sqrt{\frac{2\pi}{B}} e^{B}, \hfill T < T_\text{Q}  
            \hfill
        \end{gathered} 
    \right.
    \label{eq:1}
\end{equation}
where $T_{Q}$ is the quantum crossover temperature, $\omega_0 = \omega_\text{p} \left(1 - i^2\right)^{1/4}$ is the oscillation frequency around the bottom of the well, $\omega_\text{p} = \sqrt{{2 e I_\text{C}}/{\hbar C}}$ is the plasma frequency, $\Delta{}U(i) = 2 E_\text{j}\left[\sqrt{1 - i^2} - i\arccos{i}\right]$ is the barrier height of a potential with a localized phase particle. Here $a_\text{TA} = 4/\left(\sqrt{1+Q k_\text{B} T / \left(1.8 \Delta{}U\right)} + 1\right)^{2}$, $B = \left({\Delta U [7.2 + 8 A/Q]}\right)/\left({\hbar \omega_0}\right)$ (with a numerical parameter $A$), $Q = \omega_0 RC$. 
The probability density of switching currents was considered as $W(i)=-\pdv{P(i)}/{i}$, where the probability $P(i) = \exp{\left(-I_\text{v}^{-1} \int\limits_{0}^{i} \sfrac{1}{\tau(i')} \,\mathrm{d}i' \right)}$ is derived from the adiabatic approximation \cite{Zhou1990Sep, Pankratov2000Feb}, bearing in mind the slow change $I_\text{v} = \pdv{i}/{t}$ in bias sweep compared to the characteristic time scales of the system. 

\section{Experimental results}
\subsection{Sample fabrication and experimental set-up}
\begin{figure}[h]
    \centering
    \includegraphics[width=\linewidth, keepaspectratio]{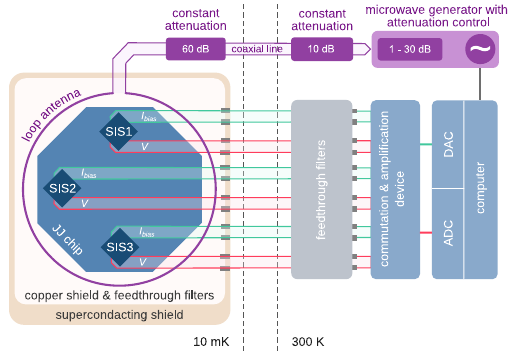}
    \caption{The experimental set-up diagram.}
    \label{fig:exp_setup}
\end{figure}

The analysis of CBJJ dynamics is carried out based on experimental data for three samples of superconductor-insulator-superconductor (SIS) tunnel junctions.
The samples of \ce{Al}/\ce{AlO_x}/\ce{Al} SIS of various areas from several squared microns to sub-micron size with  critical currents below $\qty{1}{\uA}$ were fabricated by self-aligned shadow evaporation technique \cite{Revin2020Jun,Pankratov2022May,Pankratov2022Jul}. The analysis of experimental data was carried out based on measurements of 3 different CBJJs: SIS1 with dimensions $14\times0.5\, \mu m^2$, SIS2 -- $2\times 0.4\, \mu m^2$ and SIS3 -- $2\times 1\, \mu m^2$, located on the same chip, see diagram in Fig.~\ref{fig:exp_setup}.

The CBJJ chip was thermally anchored to the mixing chamber plate of a dry dilution refrigerator, providing a minimum nominal temperature of about \qty{12}{\mK} with a temperature stability $\sim$\qty{1}{\mK} for operation below \qty{1}{\K}. The sample holder is surrounded by mu-metal and superconducting shields in addition to the copper shield on a 1K plate, decreasing background radiation \cite{Barends2011Sep}. To minimize low-frequency noise, anti-vibration dampers for the fridge body were used. Well-filtered twisted pair lines were used for conventional four-point measurements. The described experimental setup, also used for single photon detection by supplying a microwave signal via coaxial cable inside the fridge, is shown in Fig.~\ref{fig:exp_setup}.

\subsection{Switching current distributions}
The switching current distributions (SCDs) $W(I)$ were reconstructed by measuring the switching statistics of a CBJJ from a zero-voltage state to a finite voltage state in the temperature range between \qty{15}{\mK} and \qty{1}{\K}. To perform these measurements, a bias current applied to the junction was ramped up at a constant rate $I_\text{v}$ with 5000 repetitions. The constant rates for each of the samples differ according to their sizes, so for SIS1 $I_\text{v} = \qty{50}{\nA/ms}$, SIS3 -- $I_\text{v} = \qty{20}{\nA/ms}$ and SIS2 -- $I_\text{v} = \qty{5}{\nA/ms}$. 
\begin{figure}[!]
    \centering
    \includegraphics[width=\linewidth, keepaspectratio]{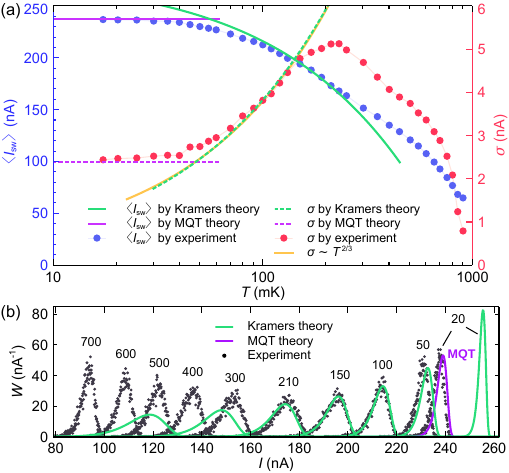}
    \caption[]{Comparison of experimental SCD for SIS3 sample with theories (\ref{eq:1}), Kramers and MQT.
    \begin{enumerate*}[label=(\alph*)]
        \item Experimental values of the mean switching current and standard deviation (blue and red circles) and theoretical estimates (solid and dashed curves).
        \item Switching current distributions $W(I)$ for various sample temperatures (marked in mK above each experimental curve). 
    \end{enumerate*}}
    \label{fig:sis3_scd}
\end{figure} 

An example of the current-voltage characteristic (IVC) for the SIS2 sample is shown in Fig.~\ref{fig:potential} c. Here the theoretical critical current (\qty{87}{\nA}, dashed line) was restored using the CBJJ parameters. Similarly, critical currents for SIS1 and SIS3 were obtained equal to \qty{895}{\nA} and \qty{270}{\nA}, respectively. From the measurements of the switching current distributions, see Fig.~\ref{fig:sis3_scd}, we extract the mean switching current $\left<I_\text{sw}\right>$ as well as the standard deviation $\sigma$ at various temperatures. Using the theory from the previous section, we reconstruct an approximation for the switching current distributions for SIS3. As can be seen from Fig.~\ref{fig:sis3_scd}, the MQT theory describes the behavior of the CBJJ where thermal noise weakly affects its dynamics. In turn, the Kramers theory is in a good agreement with experiment for TA regime. Fitting the standard deviation curves by $\sigma \sim T^{2/3}$ dependence \cite{Kurkijarvi1972Aug} is an efficient tool to separate CBJJ switching regimes instead of recoursing to detailed fitting, since this dependence (yellow solid curve) actually coincides with the Kramers theory (green dashed curve).
\begin{figure}[!]
    \centering
    \includegraphics[width=\linewidth, keepaspectratio]{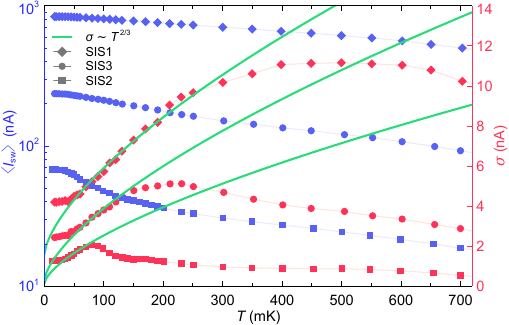}
    \caption{Experimental values of the mean switching current and standard deviation. The blue markers are the mean switching current $\langle I_\text{sw}\rangle$ and the red markers are the standard deviation $\sigma$. 
        Green solid curves show the TA mechanism $\sigma\sim T^{2/3}$ \cite{Kurkijarvi1972Aug}.
    }
    \label{fig:samples}
\end{figure} 

Qualitatively, the temperature dependence of the standard deviation is similar for all samples, see Fig.~\ref{fig:samples}, and three main switching regimes of the CBJJ are clearly visible here. At low temperatures MQT determines the dynamics of the detector. Indeed, below \qty{50}{\mK} the $\sigma (T)$ dependence is saturated, see Fig.~\ref{fig:samples}. As temperature increases, the growth of $\sigma(T)$ accelerates. This indicates the transition between MQT regime  and dominating TA dynamics. Here the dependence $\sigma(T)$ can be approximated by the function $T^{2/3}$ (see green curves in Fig.~\ref{fig:samples}), known for TA switching \cite{Kurkijarvi1972Aug}. Consequently, a separation between MQT and TA regimes (quantum crossover) is clearly visible. With further temperature increase, a deviation from TA behavior due to transition to the phase diffusion regime is observed.
In spite of different nominal critical currents (\qty{87}{\nA}, \qty{270}{\nA} and \qty{895}{\nA}) all these CBJJs demonstrate the transition to the phase diffusion regime (the point of deviation of $\sigma(T)$ from $T^{2/3}$ dependence), see Fig.~\ref{fig:samples}.

\subsection{Lifetimes}
\begin{figure*}[!]
    \centering
    \includegraphics[width=\linewidth, keepaspectratio]{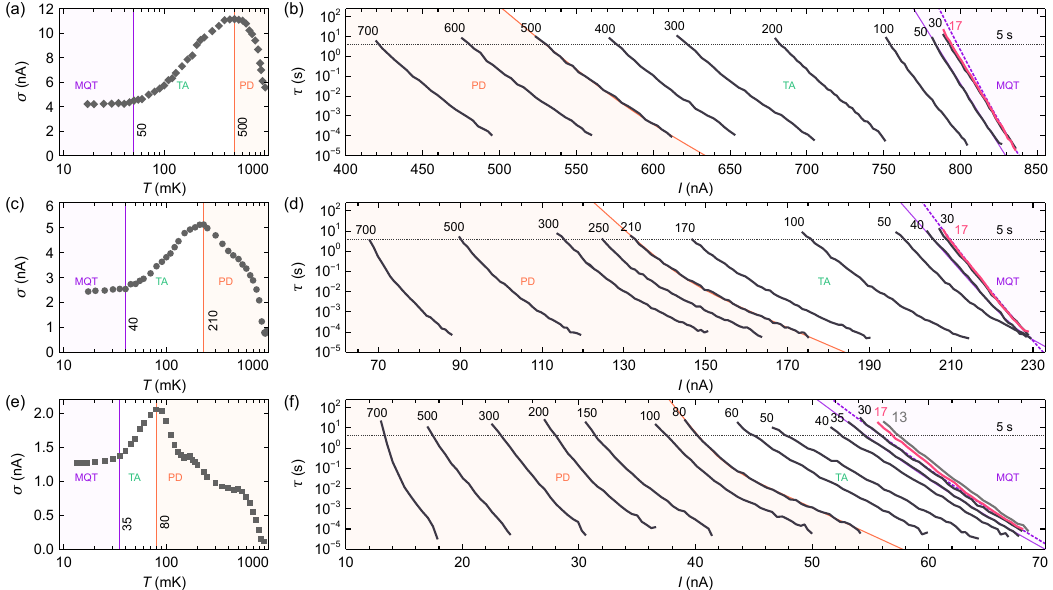}
    \caption{The experimental data at different temperatures. Standard deviation of switching currents (left) and lifetimes (right) for different sample temperatures (marked in mK above each curve).
    (a, b) -- SIS1; (c, d) -- SIS3; (e, f) -- SIS2.
    The violet area corresponds to MQT regime, the white area corresponds to TA regime, the orange area corresponds to PD regime. Violet dashed line is the MQT theory lifetime.}
    \label{fig:samples_scd_lft}
\end{figure*}
In addition to switching current distributions, an important characteristic for microwave single photon detector applications is the lifetime of zero voltage state (dark count time) $\tau(T)$.
In Fig.~\ref{fig:samples_scd_lft}b,d,f the dark count time curves $\tau(T)$ as a function of bias current and temperature (shown in mK at each curve) are presented. Here, the important characteristic of the lifetime is not only its value at a certain bias current, but also its tilt. It is obvious, that for more steeper tilts, taking the desired dark count time value, one can more closely approach the critical current, thus decreasing the detection threshold and improving the sensitivity. Usually, it is assumed that the lifetime tilt increases with decreasing temperature, which is not always so, as we show below.

By making use of experimental data of $\sigma(T)$ (Fig.~\ref{fig:samples_scd_lft}a,c,e) we present different switching regimes for lifetimes $\tau(T)$ (Fig.~\ref{fig:samples_scd_lft}b,d,f). Areas of different types of switching are highlighted: MQT (violet area), TA (white area), PD (orange area). 
Essentially, all samples behave similarly and differ in the scale of each region only.
A tendency here is visible in Fig.~\ref{fig:samples_scd_lft}: the lower the critical current of the sample, the more narrow the white region TA and the broader the orange region PD, and the MQT area is reduced as well. 

Usually, below $T_{Q}$, the lifetime can be predicted by the MQT theory \cite{Martinis1988Aug, Oelsner2013Sep, Golubev2021Jul}, but for the quantum phase diffusion regime some corrections are expected. Basically, quantum PD can manifest itself as an increase in lifetimes compared to those calculated by MQT theory. However, for CBJJs with high critical current the quantum PD effect was not demonstrated. Indeed for SIS1 ($I_C = \qty{895}{\nA}$) the experimentally observed lifetime is even shorter than predicted by theory, see Fig.~\ref{fig:samples_scd_lft}b, curve at \qty{17}{\mK}. For SIS3 sample ($I_C = \qty{270}{\nA}$) the theoretical expectation is consistent with experimental observation, see Fig.~\ref{fig:samples_scd_lft}d, curve at \qty{17}{\mK}. Visible improvement of the lifetime is observed for SIS2 only ($I_C = \qty{87}{\nA}$),  see Fig.~\ref{fig:samples_scd_lft}f, curves at \qty{17} and \qty{13}{\mK}. In spite of the fact that quantum PD regime is observed here, it does not seriously improve the detector performance. For instance, for the same lifetime of 5 s the difference between the predicted by MQT bias current and experimental one is \qty{1}{\nA} only. This is not sufficient since the photon-induced current pulse itself is of about \qty{20}{\nA}, see below.

In the TA regime the lifetimes, like the standard deviation, are traditionally described by the theory based on the Kramers expression $\tau_{TA}(T)$ (\ref{eq:1}). However, this description does not apply to the PD regime. Indeed, $\sigma (T)$ decreases with increasing temperature, reaching values even below the quantum saturation level at low temperatures, see Fig.~\ref{fig:samples_scd_lft}c,e. In parallel, the lifetime tilt $\left|{d \tau}/{dI}\right|$ increases as well, indicating that larger bias current can be used for the same lifetime $\tau$.  

\subsection{Theoretical estimates of single photon detection}
Since we have determined the mean switching currents and lifetimes as a function of temperature and bias current for our CBJJs, we can analyse their performance as single photon detectors. Assuming an ideal coupling between the CBJJ and the incident photons, following \cite{Kuzmin2018Jun}, the photon-induced  current can be estimated with account of the  CBJJ losses, described by the quality factor $Q$.
In this case the photon energy $\hbar \omega_\text{ph}$ ($\omega_\text{ph} / 2\pi$ is the photon frequency) transfers to the energy of the SIS current, therefore $I_\text{ph} = \sqrt{2 \hbar \omega_\text{ph} L^{-1}_\text{J}\left(1+2\pi/Q\right)^{-1}} $, with the Josephson inductance $L_\text{J} = \hbar/(2e \sqrt{I_\text{C}^2-I^2})$. 

As it follows from the expression above, the photon-induced current pulse amplitude differs for the investigated CBJJs since they have different parameters. Indeed, for SIS1 sample, at frequency \qty{10}{\GHz}, the magnitude of the current pulse is about \qty{50}{\nA} with a weak temperature variation. For samples with smaller critical currents the magnitude of the current pulse depends on temperature. For SIS3 this magnitude changes from \qty{25}{\nA} to \qty{35}{\nA}, for SIS2 from \qty{12}{\nA} to \qty{22}{\nA} when temperature drops from \qty{700}{\mK} to \qty{17}{\mK}.

\begin{figure}
    \centering
    \includegraphics[width=\linewidth]{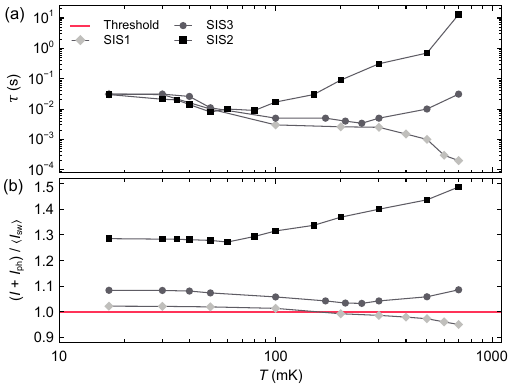}
    \caption[]{Theoretical estimate of the single photon detection efficiency.
    \begin{enumerate*}[label=(\alph*)]
        \item Lifetime with a fixed shift in bias current ($I=\langle I_\text{sw}\rangle - \Delta I$) for each sample: SIS1 -- $\Delta I$ = \qty{25}{\nA}, SIS3 -- $\Delta I$ = \qty{18}{\nA} and SIS2 -- $\Delta I$ = \qty{6}{\nA}.
        \item Normalized total current $\left(I + I_\text{ph}\right)/\langle I_\text{sw}\rangle$ due to absorption of \qty{10}{\GHz} photon versus temperature of samples. The threshold is a normalized critical current.
    \end{enumerate*}}
    \label{fig:ph_imp}
\end{figure}

To demonstrate temperature dependences of lifetimes, we fixed a bias current shift $\Delta I = \langle I_\text{sw}\rangle - I$, taking it to be sufficiently smaller than the magnitude of the photon-induced current pulse to ensure proper switching of the detector. For each sample, we take the following values: SIS1 -- $\Delta I$ = \qty{25}{\nA}, SIS3 -- $\Delta I$ = \qty{18}{\nA} and SIS2 -- $\Delta I$ = \qty{6}{\nA}. The corresponding lifetime values, subtracted from Fig.~\ref{fig:samples_scd_lft}b,d,f, are presented in Fig.~\ref{fig:ph_imp}a. 

Additionally, by choosing the desired dark count time, $\tau =5\, \rm{s}$, as shown in Fig.~\ref{fig:samples_scd_lft}b,d,f by dotted line, we can reconstruct the corresponding bias current values for each temperature. Selecting as an example the photon frequency of about \qty{10}{\GHz} \cite{Graham2024Feb}, and taking into account $I_\text{C}(T)$ dependence, we calculated the induced currents. Unambiguous switching of the detector occurs if $I + I_\text{ph} > \langle I_\text{sw}\rangle$. 
Results of our estimates are summarized in Fig.~\ref{fig:ph_imp}b. The switching threshold is indicated in this figure by the red solid line. 

The switching characteristics of CBJJ between zero and finite voltage states in the phase diffusion regime, presented in Fig.~\ref{fig:ph_imp}, are quite unusual. While for the sample with higher critical current (SIS1, $I_C = \qty{895}{\nA}$) both lifetime and total switching current $(I + I_\text{ph})/\langle I_\text{sw}\rangle$ are decreasing with temperature rise as expected, these characteristics for samples with lower critical currents exhibit non-monotonic behaviour. A decrease of the lifetime and the total switching current at temperatures slightly above the crossover temperature between the MQT and TA regimes is "compensated" by an increase of these values in the PD regime. Importantly, for samples with lower critical currents, this increase is more pronounced. This observation is counter-intuitive.
Indeed, as the critical current $I_\text{C}$ decreases, the well depth of the washboard potential becomes smaller, since $E_\text{J} \sim I_\text{C}$, see Fig.~\ref{fig:potential}a. Automatically, the ratio between the Josephson $E_\text{J}$ and the thermal energies $k_\text{B}T$ decreases. Consequently, the transition to the running (finite voltage) state requires less thermal energy, which contradicts the experiment. Qualitatively it can be explained as follows. With increase of the temperature, the probability of escaping particle re-trapping in an adjacent potential well increases. Since the threshold detector recognizes the only difference between zero and finite voltage states, it ignores noise-induced escapes, unless these escapes produce finite voltage, which improves overall detector performance. Theoretical explanation of this effect requires further study with outlining the roles of damping and temperature on the CBJJ dynamics.

\subsection{Photon detection experiments}

Based on the above-described reasoning of the possible microwave photon detection at sub-K temperatures, we have performed measurements as follows. The external microwave signal generated by the synthesizer is strongly attenuated, thus decaying into a stream of photons obeying Poisson statistics \cite{Fox2006Apr, Goltsman2001Aug, Pankratov2022May}. 

The attenuating stages include the synthesizer built-in attenuator with a power adjustment range of \qty{1}-\qty{30}{\dB}, \qty{10}{\dB} attenuator after the synthesizer output and \qty{60}{dB} attenuators distributed along the coaxial line inside the refrigerator. The signal is emitted by the loop antenna, weakly coupled to the sample (Fig. \ref{fig:exp_setup}). This technique for detecting microwave photons with power calibration using the photon-assisted tunnelling (PAT) steps at the IVC branch \cite{Tien1963Jan, Tucker1985Oct} was described in detail in \cite{Pankratov2022May}.
\begin{figure}[h!]
    \centering
    \includegraphics[width=\linewidth]{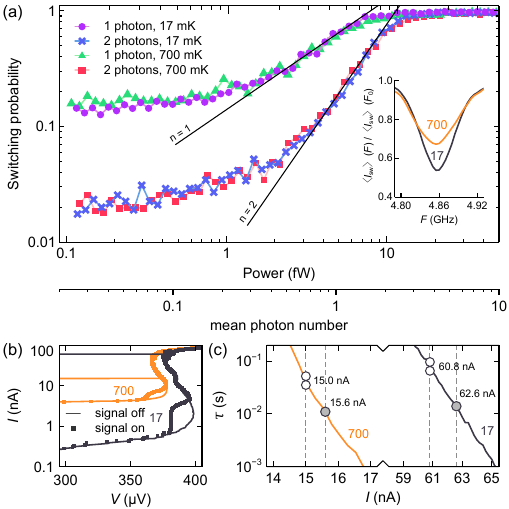}
    \caption[]{
    \begin{enumerate*}[label=(\alph*)]
        \item Switching probability versus normalized ac power of single and double \qty{4.86}{GHz} photon detection at \qty{17}{\mK} and \qty{700}{\mK} by using SIS2. The solid line n = 1 corresponds to the single photon detection tilt, and the solid line  n = 2 corresponds to the  double photon detection tilt. The inset is the renormilized average of switching current vs microwave signal frequency.
        \item The IVC for SIS2 with microwave signal (square markers, PAT steps) and without (solid curves) at temperatures \qty{17}{\mK} (black) and \qty{700}{\mK} (orange).
        \item The lifetime values corresponding to the bias currents where the detection of single and double photons occurs for temperatures of \qty{17}{\mK} and \qty{700}{\mK}. Grey markers correspond to the single photon and white markers correspond to the double photon detection.
    \end{enumerate*}}
    \label{fig:detection}
\end{figure}

After the signal has been calibrated by the PAT steps, we use the built-in attenuator of the synthesizer and collect the switching probability versus attenuated power, see Fig.~\ref{fig:detection}a, thus transferring to rare photon flux at the scale of the detector switching time, being of the order of 1 ns, while the detector dead time is restricted by the used RC filters and is about 1-10 ms. Here, due to the PAT steps, see Fig.~\ref{fig:detection}b, 20 dBm output power of the synthesizer corresponds to the accepted power of \qty{65}{fW} at \qty{17}{\mK} and \qty{48}{fW} at \qty{700}{\mK} and the curves in Fig.~\ref{fig:detection}a are normalized respectively. Therefore, the accepted power slightly decreases due to the suppressed gap. 

The microwave signal frequency is set to \qty{4.86}{GHz}. This is due to the fact that for plasma frequency close to \qty{10}{GHz} for SIS2 sample, the maximal response is expected to be at roughly half of the plasma frequency \cite{Yablokov2020Jul, Yablokov2021Jul, Ladeynov2023Jun}, that is confirmed by the SCD versus frequency measurements at a fixed output power at \qty{17}{\mK} and \qty{700}{\mK}, shown in the inset of Fig.~\ref{fig:detection}a. It should be noted that, the considered detector has rather narrow bandwidth, not exceeding 3 \% of the central frequency, decreasing, therefore, the effect of background photons.

The detection of microwave photons for two temperatures of \qty{17}{\mK} and \qty{700}{\mK} are shown in Fig.~\ref{fig:detection}a.  One can see that the switching probability curves versus power, both at \qty{17}{\mK} and \qty{700}{\mK}, have tilts corresponding to single photon and double photons detection \cite{Goltsman2001Aug} (one and two orders of probability to one order of power, respectively). Therefore, for 5 GHz microwave photons, we get about 15 \% efficiency of single photon detection even at sub-K temperatures. Important to note that the dark count floor (nearly constant part of the switching probability curves close to 0.1 level) is almost the same for both \qty{700}{\mK} and \qty{17}{\mK} temperatures. The observed weak temperature dependence of the detector performance is fully consistent with the analysis presented above.

\section{Conclusions}
We have experimentally demonstrated single and double photons sensing at $\approx$\qty{5}{GHz} by making use of a threshold detector, based on \ce{Al}-made Josephson junctions with relatively low critical currents, exhibiting a phase diffusion regime. For two-photon detection, the obtained characteristics of the detector are quite impressive - with a dark count time $\tau$ of 0.1 s its efficiency is 90 \%. However, for a single photon detection the obtained parameters are not so attractive: $\tau \approx 0.01$ s and the efficiency is 15 \%. Nevertheless, it is important to note that for a particular sample with dominating phase diffusion regime these numbers are almost temperature independent. This means that the main mechanism limiting detector performance is "unwanted" photons rather than thermal fluctuations. Further research is required to characterize a source of such uncontrolled photons.  These results are important for axion search experiments in the range of 5-25 GHz (20-100 $\mu$eV) \cite{Buschmann2022Feb}. Also, 
operation of the detector at 0.7 K allows one to move the detector from a dilution fridge plate to a condensing 0.7 K plate. This allows increasing a distance from a strong magnet, used for axion search and usually located just below the dilution plate, thus reducing the effect of a high magnetic field on the detector.

\begin{acknowledgments}
The work is supported by the Russian Science Foundation (Project~\mbox{19-79-10170}).
\end{acknowledgments}

% \bibliography{main}

\begin{thebibliography}{72}%
\makeatletter
\providecommand \@ifxundefined [1]{%
 \@ifx{#1\undefined}
}%
\providecommand \@ifnum [1]{%
 \ifnum #1\expandafter \@firstoftwo
 \else \expandafter \@secondoftwo
 \fi
}%
\providecommand \@ifx [1]{%
 \ifx #1\expandafter \@firstoftwo
 \else \expandafter \@secondoftwo
 \fi
}%
\providecommand \natexlab [1]{#1}%
\providecommand \enquote  [1]{``#1''}%
\providecommand \bibnamefont  [1]{#1}%
\providecommand \bibfnamefont [1]{#1}%
\providecommand \citenamefont [1]{#1}%
\providecommand \href@noop [0]{\@secondoftwo}%
\providecommand \href [0]{\begingroup \@sanitize@url \@href}%
\providecommand \@href[1]{\@@startlink{#1}\@@href}%
\providecommand \@@href[1]{\endgroup#1\@@endlink}%
\providecommand \@sanitize@url [0]{\catcode `\\12\catcode `\$12\catcode
  `\&12\catcode `\#12\catcode `\^12\catcode `\_12\catcode `\%12\relax}%
\providecommand \@@startlink[1]{}%
\providecommand \@@endlink[0]{}%
\providecommand \url  [0]{\begingroup\@sanitize@url \@url }%
\providecommand \@url [1]{\endgroup\@href {#1}{\urlprefix }}%
\providecommand \urlprefix  [0]{URL }%
\providecommand \Eprint [0]{\href }%
\providecommand \doibase [0]{https://doi.org/}%
\providecommand \selectlanguage [0]{\@gobble}%
\providecommand \bibinfo  [0]{\@secondoftwo}%
\providecommand \bibfield  [0]{\@secondoftwo}%
\providecommand \translation [1]{[#1]}%
\providecommand \BibitemOpen [0]{}%
\providecommand \bibitemStop [0]{}%
\providecommand \bibitemNoStop [0]{.\EOS\space}%
\providecommand \EOS [0]{\spacefactor3000\relax}%
\providecommand \BibitemShut  [1]{\csname bibitem#1\endcsname}%
\let\auto@bib@innerbib\@empty
%</preamble>
\bibitem [{\citenamefont {Esmaeil~Zadeh}\ \emph {et~al.}(2021)\citenamefont
  {Esmaeil~Zadeh}, \citenamefont {Chang}, \citenamefont {Los}, \citenamefont
  {Gyger}, \citenamefont {Elshaari}, \citenamefont {Steinhauer}, \citenamefont
  {Dorenbos},\ and\ \citenamefont {Zwiller}}]{EsmaeilZadeh2021May}%
  \BibitemOpen
  \bibfield  {author} {\bibinfo {author} {\bibfnamefont {I.}~\bibnamefont
  {Esmaeil~Zadeh}}, \bibinfo {author} {\bibfnamefont {J.}~\bibnamefont
  {Chang}}, \bibinfo {author} {\bibfnamefont {J.~W.~N.}\ \bibnamefont {Los}},
  \bibinfo {author} {\bibfnamefont {S.}~\bibnamefont {Gyger}}, \bibinfo
  {author} {\bibfnamefont {A.~W.}\ \bibnamefont {Elshaari}}, \bibinfo {author}
  {\bibfnamefont {S.}~\bibnamefont {Steinhauer}}, \bibinfo {author}
  {\bibfnamefont {S.~N.}\ \bibnamefont {Dorenbos}},\ and\ \bibinfo {author}
  {\bibfnamefont {V.}~\bibnamefont {Zwiller}},\ }\bibfield  {title} {\bibinfo
  {title} {{Superconducting nanowire single-photon detectors: A perspective on
  evolution, state-of-the-art, future developments, and applications}},\
  }\bibfield  {journal} {\bibinfo  {journal} {Applied Physics Letters}\
  }\textbf {\bibinfo {volume} {118}},\ \href
  {https://doi.org/10.1063/5.0045990} (\bibinfo {year}
  {2021})\BibitemShut {NoStop}%
\bibitem [{\citenamefont {Khasminskaya}\ \emph {et~al.}(2016)\citenamefont
  {Khasminskaya}, \citenamefont {Pyatkov}, \citenamefont {Słowik},
  \citenamefont {Ferrari}, \citenamefont {Kahl}, \citenamefont {Kovalyuk},
  \citenamefont {Rath}, \citenamefont {Vetter}, \citenamefont {Hennrich},
  \citenamefont {Kappes}, \citenamefont {Gol'tsman}, \citenamefont {Korneev},
  \citenamefont {Rockstuhl}, \citenamefont {Krupke},\ and\ \citenamefont
  {Pernice}}]{Khasminskaya2016Nov}%
  \BibitemOpen
  \bibfield  {author} {\bibinfo {author} {\bibfnamefont {S.}~\bibnamefont
  {Khasminskaya}}, \bibinfo {author} {\bibfnamefont {F.}~\bibnamefont
  {Pyatkov}}, \bibinfo {author} {\bibfnamefont {K.}~\bibnamefont {Słowik}},
  \bibinfo {author} {\bibfnamefont {S.}~\bibnamefont {Ferrari}}, \bibinfo
  {author} {\bibfnamefont {O.}~\bibnamefont {Kahl}}, \bibinfo {author}
  {\bibfnamefont {V.}~\bibnamefont {Kovalyuk}}, \bibinfo {author}
  {\bibfnamefont {P.}~\bibnamefont {Rath}}, \bibinfo {author} {\bibfnamefont
  {A.}~\bibnamefont {Vetter}}, \bibinfo {author} {\bibfnamefont
  {F.}~\bibnamefont {Hennrich}}, \bibinfo {author} {\bibfnamefont {M.~M.}\
  \bibnamefont {Kappes}}, \bibinfo {author} {\bibfnamefont {G.}~\bibnamefont
  {Gol'tsman}}, \bibinfo {author} {\bibfnamefont {A.}~\bibnamefont {Korneev}},
  \bibinfo {author} {\bibfnamefont {C.}~\bibnamefont {Rockstuhl}}, \bibinfo
  {author} {\bibfnamefont {R.}~\bibnamefont {Krupke}},\ and\ \bibinfo {author}
  {\bibfnamefont {W.~H.~P.}\ \bibnamefont {Pernice}},\ }\bibfield  {title}
  {\bibinfo {title} {{Fully integrated quantum photonic circuit with an
  electrically driven light source}},\ }\href
  {https://doi.org/10.1038/nphoton.2016.178} {\bibfield  {journal} {\bibinfo
  {journal} {Nature Photonics}\ }\textbf {\bibinfo {volume} {10}},\ \bibinfo
  {pages} {727} (\bibinfo {year} {2016})}\BibitemShut {NoStop}%
\bibitem [{\citenamefont {Kahl}\ \emph {et~al.}(2015)\citenamefont {Kahl},
  \citenamefont {Ferrari}, \citenamefont {Kovalyuk}, \citenamefont {Goltsman},
  \citenamefont {Korneev},\ and\ \citenamefont {Pernice}}]{Kahl2015Jun}%
  \BibitemOpen
  \bibfield  {author} {\bibinfo {author} {\bibfnamefont {O.}~\bibnamefont
  {Kahl}}, \bibinfo {author} {\bibfnamefont {S.}~\bibnamefont {Ferrari}},
  \bibinfo {author} {\bibfnamefont {V.}~\bibnamefont {Kovalyuk}}, \bibinfo
  {author} {\bibfnamefont {G.~N.}\ \bibnamefont {Goltsman}}, \bibinfo {author}
  {\bibfnamefont {A.}~\bibnamefont {Korneev}},\ and\ \bibinfo {author}
  {\bibfnamefont {W.~H.~P.}\ \bibnamefont {Pernice}},\ }\bibfield  {title}
  {\bibinfo {title} {{Waveguide integrated superconducting single-photon
  detectors with high internal quantum efficiency at telecom wavelengths}},\
  }\href {https://doi.org/10.1038/srep10941} {\bibfield  {journal} {\bibinfo
  {journal} {Sci. Rep.}\ }\textbf {\bibinfo {volume} {5}},\ \bibinfo {pages}
  {10941} (\bibinfo {year} {2015})}\BibitemShut {NoStop}%
\bibitem [{\citenamefont {Irwin}\ and\ \citenamefont
  {Hilton}(2005)}]{Irwin2005Jul}%
  \BibitemOpen
  \bibfield  {author} {\bibinfo {author} {\bibfnamefont {K.~D.}\ \bibnamefont
  {Irwin}}\ and\ \bibinfo {author} {\bibfnamefont {G.~C.}\ \bibnamefont
  {Hilton}},\ }\bibinfo {title} {{Transition-Edge Sensors}},\ in\ \href
  {https://doi.org/10.1007/10933596_3} {\emph {\bibinfo {booktitle} {{Cryogenic
  Particle Detection}}}}\ (\bibinfo  {publisher} {Springer},\ \bibinfo
  {address} {Berlin, Heidelberg, Germany},\ \bibinfo {year} {2005})\ pp.\
  \bibinfo {pages} {63--150}\BibitemShut {NoStop}%
\bibitem [{\citenamefont {F\"{o}rtsch}\ \emph {et~al.}(2015)\citenamefont
  {F\"{o}rtsch}, \citenamefont {Gerrits}, \citenamefont {Stevens},
  \citenamefont {Strekalov}, \citenamefont {Schunk}, \citenamefont {F\"{u}rst},
  \citenamefont {Vogl}, \citenamefont {Sedlmeir}, \citenamefont {Schwefel},
  \citenamefont {Leuchs}, \citenamefont {Nam},\ and\ \citenamefont
  {Marquardt}}]{Fortsch2015Apr}%
  \BibitemOpen
  \bibfield  {author} {\bibinfo {author} {\bibfnamefont {M.}~\bibnamefont
  {F\"{o}rtsch}}, \bibinfo {author} {\bibfnamefont {T.}~\bibnamefont
  {Gerrits}}, \bibinfo {author} {\bibfnamefont {M.~J.}\ \bibnamefont
  {Stevens}}, \bibinfo {author} {\bibfnamefont {D.}~\bibnamefont {Strekalov}},
  \bibinfo {author} {\bibfnamefont {G.}~\bibnamefont {Schunk}}, \bibinfo
  {author} {\bibfnamefont {J.~U.}\ \bibnamefont {F\"{u}rst}}, \bibinfo {author}
  {\bibfnamefont {U.}~\bibnamefont {Vogl}}, \bibinfo {author} {\bibfnamefont
  {F.}~\bibnamefont {Sedlmeir}}, \bibinfo {author} {\bibfnamefont {H.~G.~L.}\
  \bibnamefont {Schwefel}}, \bibinfo {author} {\bibfnamefont {G.}~\bibnamefont
  {Leuchs}}, \bibinfo {author} {\bibfnamefont {S.~W.}\ \bibnamefont {Nam}},\
  and\ \bibinfo {author} {\bibfnamefont {C.}~\bibnamefont {Marquardt}},\
  }\bibfield  {title} {\bibinfo {title} {{Near-infrared single-photon
  spectroscopy of a whispering gallery mode resonator using energy-resolving
  transition edge sensors}},\ }\href
  {https://doi.org/10.1088/2040-8978/17/6/065501} {\bibfield  {journal}
  {\bibinfo  {journal} {J. Opt.}\ }\textbf {\bibinfo {volume} {17}},\ \bibinfo
  {pages} {065501} (\bibinfo {year} {2015})}\BibitemShut {NoStop}%
\bibitem [{\citenamefont {Johnson}\ \emph {et~al.}(2010)\citenamefont
  {Johnson}, \citenamefont {Reed}, \citenamefont {Houck}, \citenamefont
  {Schuster}, \citenamefont {Bishop}, \citenamefont {Ginossar}, \citenamefont
  {Gambetta}, \citenamefont {DiCarlo}, \citenamefont {Frunzio}, \citenamefont
  {Girvin},\ and\ \citenamefont {Schoelkopf}}]{Johnson2010Sep}%
  \BibitemOpen
  \bibfield  {author} {\bibinfo {author} {\bibfnamefont {B.~R.}\ \bibnamefont
  {Johnson}}, \bibinfo {author} {\bibfnamefont {M.~D.}\ \bibnamefont {Reed}},
  \bibinfo {author} {\bibfnamefont {A.~A.}\ \bibnamefont {Houck}}, \bibinfo
  {author} {\bibfnamefont {D.~I.}\ \bibnamefont {Schuster}}, \bibinfo {author}
  {\bibfnamefont {L.~S.}\ \bibnamefont {Bishop}}, \bibinfo {author}
  {\bibfnamefont {E.}~\bibnamefont {Ginossar}}, \bibinfo {author}
  {\bibfnamefont {J.~M.}\ \bibnamefont {Gambetta}}, \bibinfo {author}
  {\bibfnamefont {L.}~\bibnamefont {DiCarlo}}, \bibinfo {author} {\bibfnamefont
  {L.}~\bibnamefont {Frunzio}}, \bibinfo {author} {\bibfnamefont {S.~M.}\
  \bibnamefont {Girvin}},\ and\ \bibinfo {author} {\bibfnamefont {R.~J.}\
  \bibnamefont {Schoelkopf}},\ }\bibfield  {title} {\bibinfo {title} {{Quantum
  non-demolition detection of single microwave photons in a circuit}},\ }\href
  {https://doi.org/10.1038/nphys1710} {\bibfield  {journal} {\bibinfo
  {journal} {Nature Physics}\ }\textbf {\bibinfo {volume} {6}},\ \bibinfo
  {pages} {663} (\bibinfo {year} {2010})}\BibitemShut {NoStop}%
\bibitem [{\citenamefont {Inomata}\ \emph {et~al.}(2016)\citenamefont
  {Inomata}, \citenamefont {Lin}, \citenamefont {Koshino}, \citenamefont
  {Oliver}, \citenamefont {Tsai}, \citenamefont {Yamamoto},\ and\ \citenamefont
  {Nakamura}}]{Inomata2016Jul}%
  \BibitemOpen
  \bibfield  {author} {\bibinfo {author} {\bibfnamefont {K.}~\bibnamefont
  {Inomata}}, \bibinfo {author} {\bibfnamefont {Z.}~\bibnamefont {Lin}},
  \bibinfo {author} {\bibfnamefont {K.}~\bibnamefont {Koshino}}, \bibinfo
  {author} {\bibfnamefont {W.~D.}\ \bibnamefont {Oliver}}, \bibinfo {author}
  {\bibfnamefont {J.-S.}\ \bibnamefont {Tsai}}, \bibinfo {author}
  {\bibfnamefont {T.}~\bibnamefont {Yamamoto}},\ and\ \bibinfo {author}
  {\bibfnamefont {Y.}~\bibnamefont {Nakamura}},\ }\bibfield  {title} {\bibinfo
  {title} {{Single microwave-photon detector using an artificial
  {$\Lambda$}-type three-level system}},\ }\href
  {https://doi.org/10.1038/ncomms12303} {\bibfield  {journal} {\bibinfo
  {journal} {Nature Communications}\ }\textbf {\bibinfo {volume} {7}},\
  \bibinfo {pages} {1} (\bibinfo {year} {2016})}\BibitemShut {NoStop}%
\bibitem [{\citenamefont {Wong}\ and\ \citenamefont
  {Vavilov}(2017)}]{Wong2017Jan}%
  \BibitemOpen
  \bibfield  {author} {\bibinfo {author} {\bibfnamefont {C.~H.}\ \bibnamefont
  {Wong}}\ and\ \bibinfo {author} {\bibfnamefont {M.~G.}\ \bibnamefont
  {Vavilov}},\ }\bibfield  {title} {\bibinfo {title} {{Quantum efficiency of a
  single microwave photon detector based on a semiconductor double quantum
  dot}},\ }\href {https://doi.org/10.1103/PhysRevA.95.012325} {\bibfield
  {journal} {\bibinfo  {journal} {Phys. Rev. A}\ }\textbf {\bibinfo {volume}
  {95}},\ \bibinfo {pages} {012325} (\bibinfo {year} {2017})}\BibitemShut
  {NoStop}%
\bibitem [{\citenamefont {Besse}\ \emph {et~al.}(2018)\citenamefont {Besse},
  \citenamefont {Gasparinetti}, \citenamefont {Collodo}, \citenamefont
  {Walter}, \citenamefont {Kurpiers}, \citenamefont {Pechal}, \citenamefont
  {Eichler},\ and\ \citenamefont {Wallraff}}]{Besse2018Apr}%
  \BibitemOpen
  \bibfield  {author} {\bibinfo {author} {\bibfnamefont {J.-C.}\ \bibnamefont
  {Besse}}, \bibinfo {author} {\bibfnamefont {S.}~\bibnamefont {Gasparinetti}},
  \bibinfo {author} {\bibfnamefont {M.~C.}\ \bibnamefont {Collodo}}, \bibinfo
  {author} {\bibfnamefont {T.}~\bibnamefont {Walter}}, \bibinfo {author}
  {\bibfnamefont {P.}~\bibnamefont {Kurpiers}}, \bibinfo {author}
  {\bibfnamefont {M.}~\bibnamefont {Pechal}}, \bibinfo {author} {\bibfnamefont
  {C.}~\bibnamefont {Eichler}},\ and\ \bibinfo {author} {\bibfnamefont
  {A.}~\bibnamefont {Wallraff}},\ }\bibfield  {title} {\bibinfo {title}
  {{Single-Shot Quantum Nondemolition Detection of Individual Itinerant
  Microwave Photons}},\ }\href {https://doi.org/10.1103/PhysRevX.8.021003}
  {\bibfield  {journal} {\bibinfo  {journal} {Phys. Rev. X}\ }\textbf {\bibinfo
  {volume} {8}},\ \bibinfo {pages} {021003} (\bibinfo {year}
  {2018})}\BibitemShut {NoStop}%
\bibitem [{\citenamefont {Royer}\ \emph {et~al.}(2018)\citenamefont {Royer},
  \citenamefont {Grimsmo}, \citenamefont {Choquette-Poitevin},\ and\
  \citenamefont {Blais}}]{Royer2018May}%
  \BibitemOpen
  \bibfield  {author} {\bibinfo {author} {\bibfnamefont {B.}~\bibnamefont
  {Royer}}, \bibinfo {author} {\bibfnamefont {A.~L.}\ \bibnamefont {Grimsmo}},
  \bibinfo {author} {\bibfnamefont {A.}~\bibnamefont {Choquette-Poitevin}},\
  and\ \bibinfo {author} {\bibfnamefont {A.}~\bibnamefont {Blais}},\ }\bibfield
   {title} {\bibinfo {title} {{Itinerant Microwave Photon Detector}},\ }\href
  {https://doi.org/10.1103/PhysRevLett.120.203602} {\bibfield  {journal}
  {\bibinfo  {journal} {Phys. Rev. Lett.}\ }\textbf {\bibinfo {volume} {120}},\
  \bibinfo {pages} {203602} (\bibinfo {year} {2018})}\BibitemShut {NoStop}%
\bibitem [{\citenamefont {Kono}\ \emph {et~al.}(2018)\citenamefont {Kono},
  \citenamefont {Koshino}, \citenamefont {Tabuchi}, \citenamefont {Noguchi},\
  and\ \citenamefont {Nakamura}}]{Kono2018Jun}%
  \BibitemOpen
  \bibfield  {author} {\bibinfo {author} {\bibfnamefont {S.}~\bibnamefont
  {Kono}}, \bibinfo {author} {\bibfnamefont {K.}~\bibnamefont {Koshino}},
  \bibinfo {author} {\bibfnamefont {Y.}~\bibnamefont {Tabuchi}}, \bibinfo
  {author} {\bibfnamefont {A.}~\bibnamefont {Noguchi}},\ and\ \bibinfo {author}
  {\bibfnamefont {Y.}~\bibnamefont {Nakamura}},\ }\bibfield  {title} {\bibinfo
  {title} {{Quantum non-demolition detection of an itinerant microwave
  photon}},\ }\href {https://doi.org/10.1038/s41567-018-0066-3} {\bibfield
  {journal} {\bibinfo  {journal} {Nature Physics}\ }\textbf {\bibinfo {volume}
  {14}},\ \bibinfo {pages} {546} (\bibinfo {year} {2018})}\BibitemShut
  {NoStop}%
\bibitem [{\citenamefont {Lescanne}\ \emph {et~al.}(2020)\citenamefont
  {Lescanne}, \citenamefont {Deléglise}, \citenamefont {Albertinale},
  \citenamefont {Réglade}, \citenamefont {Capelle}, \citenamefont {Ivanov},
  \citenamefont {Jacqmin}, \citenamefont {Leghtas},\ and\ \citenamefont
  {Flurin}}]{Lescanne2020May}%
  \BibitemOpen
  \bibfield  {author} {\bibinfo {author} {\bibfnamefont {R.}~\bibnamefont
  {Lescanne}}, \bibinfo {author} {\bibfnamefont {S.}~\bibnamefont
  {Deléglise}}, \bibinfo {author} {\bibfnamefont {E.}~\bibnamefont
  {Albertinale}}, \bibinfo {author} {\bibfnamefont {U.}~\bibnamefont
  {Réglade}}, \bibinfo {author} {\bibfnamefont {T.}~\bibnamefont {Capelle}},
  \bibinfo {author} {\bibfnamefont {E.}~\bibnamefont {Ivanov}}, \bibinfo
  {author} {\bibfnamefont {T.}~\bibnamefont {Jacqmin}}, \bibinfo {author}
  {\bibfnamefont {Z.}~\bibnamefont {Leghtas}},\ and\ \bibinfo {author}
  {\bibfnamefont {E.}~\bibnamefont {Flurin}},\ }\bibfield  {title} {\bibinfo
  {title} {{Irreversible Qubit-Photon Coupling for the Detection of Itinerant
  Microwave Photons}},\ }\href {https://doi.org/10.1103/PhysRevX.10.021038}
  {\bibfield  {journal} {\bibinfo  {journal} {Phys. Rev. X}\ }\textbf {\bibinfo
  {volume} {10}},\ \bibinfo {pages} {021038} (\bibinfo {year}
  {2020})}\BibitemShut {NoStop}%
\bibitem [{\citenamefont {Albertinale}\ \emph {et~al.}(2021)\citenamefont
  {Albertinale}, \citenamefont {Balembois}, \citenamefont {Billaud},
  \citenamefont {Ranjan}, \citenamefont {Flanigan}, \citenamefont {Schenkel},
  \citenamefont {Est{\`e}ve}, \citenamefont {Vion}, \citenamefont {Bertet},\
  and\ \citenamefont {Flurin}}]{Albertinale2021}%
  \BibitemOpen
  \bibfield  {author} {\bibinfo {author} {\bibfnamefont {E.}~\bibnamefont
  {Albertinale}}, \bibinfo {author} {\bibfnamefont {L.}~\bibnamefont
  {Balembois}}, \bibinfo {author} {\bibfnamefont {E.}~\bibnamefont {Billaud}},
  \bibinfo {author} {\bibfnamefont {V.}~\bibnamefont {Ranjan}}, \bibinfo
  {author} {\bibfnamefont {D.}~\bibnamefont {Flanigan}}, \bibinfo {author}
  {\bibfnamefont {T.}~\bibnamefont {Schenkel}}, \bibinfo {author}
  {\bibfnamefont {D.}~\bibnamefont {Est{\`e}ve}}, \bibinfo {author}
  {\bibfnamefont {D.}~\bibnamefont {Vion}}, \bibinfo {author} {\bibfnamefont
  {P.}~\bibnamefont {Bertet}},\ and\ \bibinfo {author} {\bibfnamefont
  {E.}~\bibnamefont {Flurin}},\ }\bibfield  {title} {\bibinfo {title}
  {Detecting spins by their fluorescence with a microwave photon counter},\
  }\href {https://doi.org/10.1038/s41586-021-04076-z} {\bibfield  {journal}
  {\bibinfo  {journal} {Nature}\ }\textbf {\bibinfo {volume} {600}},\ \bibinfo
  {pages} {434} (\bibinfo {year} {2021})}\BibitemShut {NoStop}%
\bibitem [{\citenamefont {Grimsmo}\ \emph {et~al.}(2021)\citenamefont
  {Grimsmo}, \citenamefont {Royer}, \citenamefont {Kreikebaum}, \citenamefont
  {Ye}, \citenamefont {O{'}Brien}, \citenamefont {Siddiqi},\ and\ \citenamefont
  {Blais}}]{Grimsmo2021Mar}%
  \BibitemOpen
  \bibfield  {author} {\bibinfo {author} {\bibfnamefont {A.~L.}\ \bibnamefont
  {Grimsmo}}, \bibinfo {author} {\bibfnamefont {B.}~\bibnamefont {Royer}},
  \bibinfo {author} {\bibfnamefont {J.~M.}\ \bibnamefont {Kreikebaum}},
  \bibinfo {author} {\bibfnamefont {Y.}~\bibnamefont {Ye}}, \bibinfo {author}
  {\bibfnamefont {K.}~\bibnamefont {O{'}Brien}}, \bibinfo {author}
  {\bibfnamefont {I.}~\bibnamefont {Siddiqi}},\ and\ \bibinfo {author}
  {\bibfnamefont {A.}~\bibnamefont {Blais}},\ }\bibfield  {title} {\bibinfo
  {title} {{Quantum Metamaterial for Broadband Detection of Single Microwave
  Photons}},\ }\href {https://doi.org/10.1103/PhysRevApplied.15.034074}
  {\bibfield  {journal} {\bibinfo  {journal} {Phys. Rev. Appl.}\ }\textbf
  {\bibinfo {volume} {15}},\ \bibinfo {pages} {034074} (\bibinfo {year}
  {2021})}\BibitemShut {NoStop}%
\bibitem [{\citenamefont {Dixit}\ \emph {et~al.}(2021)\citenamefont {Dixit},
  \citenamefont {Chakram}, \citenamefont {He}, \citenamefont {Agrawal},
  \citenamefont {Naik}, \citenamefont {Schuster},\ and\ \citenamefont
  {Chou}}]{Dixit2021Apr}%
  \BibitemOpen
  \bibfield  {author} {\bibinfo {author} {\bibfnamefont {A.~V.}\ \bibnamefont
  {Dixit}}, \bibinfo {author} {\bibfnamefont {S.}~\bibnamefont {Chakram}},
  \bibinfo {author} {\bibfnamefont {K.}~\bibnamefont {He}}, \bibinfo {author}
  {\bibfnamefont {A.}~\bibnamefont {Agrawal}}, \bibinfo {author} {\bibfnamefont
  {R.~K.}\ \bibnamefont {Naik}}, \bibinfo {author} {\bibfnamefont {D.~I.}\
  \bibnamefont {Schuster}},\ and\ \bibinfo {author} {\bibfnamefont
  {A.}~\bibnamefont {Chou}},\ }\bibfield  {title} {\bibinfo {title} {{Searching
  for Dark Matter with a Superconducting Qubit}},\ }\href
  {https://doi.org/10.1103/PhysRevLett.126.141302} {\bibfield  {journal}
  {\bibinfo  {journal} {Phys. Rev. Lett.}\ }\textbf {\bibinfo {volume} {126}},\
  \bibinfo {pages} {141302} (\bibinfo {year} {2021})}\BibitemShut {NoStop}%
\bibitem [{\citenamefont {Khan}\ \emph {et~al.}(2021)\citenamefont {Khan},
  \citenamefont {Potts}, \citenamefont {Lehmann}, \citenamefont {Thelander},
  \citenamefont {Dick}, \citenamefont {Samuelsson},\ and\ \citenamefont
  {Maisi}}]{Khan2021Aug}%
  \BibitemOpen
  \bibfield  {author} {\bibinfo {author} {\bibfnamefont {W.}~\bibnamefont
  {Khan}}, \bibinfo {author} {\bibfnamefont {P.~P.}\ \bibnamefont {Potts}},
  \bibinfo {author} {\bibfnamefont {S.}~\bibnamefont {Lehmann}}, \bibinfo
  {author} {\bibfnamefont {C.}~\bibnamefont {Thelander}}, \bibinfo {author}
  {\bibfnamefont {K.~A.}\ \bibnamefont {Dick}}, \bibinfo {author}
  {\bibfnamefont {P.}~\bibnamefont {Samuelsson}},\ and\ \bibinfo {author}
  {\bibfnamefont {V.~F.}\ \bibnamefont {Maisi}},\ }\bibfield  {title} {\bibinfo
  {title} {{Efficient and continuous microwave photoconversion in hybrid
  cavity-semiconductor nanowire double quantum dot diodes}},\ }\href
  {https://doi.org/10.1038/s41467-021-25446-1} {\bibfield  {journal} {\bibinfo
  {journal} {Nat. Commun.}\ }\textbf {\bibinfo {volume} {12}},\ \bibinfo
  {pages} {1} (\bibinfo {year} {2021})}\BibitemShut {NoStop}%
\bibitem [{\citenamefont {Balembois}\ \emph {et~al.}(2024)\citenamefont
  {Balembois}, \citenamefont {Travesedo}, \citenamefont {Pallegoix},
  \citenamefont {May}, \citenamefont {Billaud}, \citenamefont {Villiers},
  \citenamefont {Est\`eve}, \citenamefont {Vion}, \citenamefont {Bertet},\ and\
  \citenamefont {Flurin}}]{Balembois2024Jan}%
  \BibitemOpen
  \bibfield  {author} {\bibinfo {author} {\bibfnamefont {L.}~\bibnamefont
  {Balembois}}, \bibinfo {author} {\bibfnamefont {J.}~\bibnamefont
  {Travesedo}}, \bibinfo {author} {\bibfnamefont {L.}~\bibnamefont
  {Pallegoix}}, \bibinfo {author} {\bibfnamefont {A.}~\bibnamefont {May}},
  \bibinfo {author} {\bibfnamefont {E.}~\bibnamefont {Billaud}}, \bibinfo
  {author} {\bibfnamefont {M.}~\bibnamefont {Villiers}}, \bibinfo {author}
  {\bibfnamefont {D.}~\bibnamefont {Est\`eve}}, \bibinfo {author}
  {\bibfnamefont {D.}~\bibnamefont {Vion}}, \bibinfo {author} {\bibfnamefont
  {P.}~\bibnamefont {Bertet}},\ and\ \bibinfo {author} {\bibfnamefont
  {E.}~\bibnamefont {Flurin}},\ }\bibfield  {title} {\bibinfo {title}
  {Cyclically operated microwave single-photon counter with sensitivity of
  ${10}^{\ensuremath{-}22}\phantom{\rule{0.2em}{0ex}}\mathrm{W}/\sqrt{\mathrm{hz}}$},\
  }\href {https://doi.org/10.1103/PhysRevApplied.21.014043} {\bibfield
  {journal} {\bibinfo  {journal} {Phys. Rev. Appl.}\ }\textbf {\bibinfo
  {volume} {21}},\ \bibinfo {pages} {014043} (\bibinfo {year}
  {2024})}\BibitemShut {NoStop}%
\bibitem [{\citenamefont {Haldar}\ \emph {et~al.}(2024)\citenamefont {Haldar},
  \citenamefont {Zenelaj}, \citenamefont {Potts}, \citenamefont {Havir},
  \citenamefont {Lehmann}, \citenamefont {Dick}, \citenamefont {Samuelsson},\
  and\ \citenamefont {Maisi}}]{Haldar2024Feb}%
  \BibitemOpen
  \bibfield  {author} {\bibinfo {author} {\bibfnamefont {S.}~\bibnamefont
  {Haldar}}, \bibinfo {author} {\bibfnamefont {D.}~\bibnamefont {Zenelaj}},
  \bibinfo {author} {\bibfnamefont {P.~P.}\ \bibnamefont {Potts}}, \bibinfo
  {author} {\bibfnamefont {H.}~\bibnamefont {Havir}}, \bibinfo {author}
  {\bibfnamefont {S.}~\bibnamefont {Lehmann}}, \bibinfo {author} {\bibfnamefont
  {K.~A.}\ \bibnamefont {Dick}}, \bibinfo {author} {\bibfnamefont
  {P.}~\bibnamefont {Samuelsson}},\ and\ \bibinfo {author} {\bibfnamefont
  {V.~F.}\ \bibnamefont {Maisi}},\ }\bibfield  {title} {\bibinfo {title}
  {{Microwave power harvesting using resonator-coupled double quantum dot
  photodiode}},\ }\href {https://doi.org/10.1103/PhysRevB.109.L081403}
  {\bibfield  {journal} {\bibinfo  {journal} {Phys. Rev. B}\ }\textbf {\bibinfo
  {volume} {109}},\ \bibinfo {pages} {L081403} (\bibinfo {year}
  {2024})}\BibitemShut {NoStop}%
\bibitem [{\citenamefont {Martinis}\ \emph {et~al.}(2002)\citenamefont
  {Martinis}, \citenamefont {Nam}, \citenamefont {Aumentado},\ and\
  \citenamefont {Urbina}}]{Martinis2002Aug}%
  \BibitemOpen
  \bibfield  {author} {\bibinfo {author} {\bibfnamefont {J.~M.}\ \bibnamefont
  {Martinis}}, \bibinfo {author} {\bibfnamefont {S.}~\bibnamefont {Nam}},
  \bibinfo {author} {\bibfnamefont {J.}~\bibnamefont {Aumentado}},\ and\
  \bibinfo {author} {\bibfnamefont {C.}~\bibnamefont {Urbina}},\ }\bibfield
  {title} {\bibinfo {title} {{Rabi Oscillations in a Large Josephson-Junction
  Qubit}},\ }\href {https://doi.org/10.1103/PhysRevLett.89.117901} {\bibfield
  {journal} {\bibinfo  {journal} {Phys. Rev. Lett.}\ }\textbf {\bibinfo
  {volume} {89}},\ \bibinfo {pages} {117901} (\bibinfo {year}
  {2002})}\BibitemShut {NoStop}%
\bibitem [{\citenamefont {Barone}\ and\ \citenamefont
  {Patern\`{o}}(1982)}]{Barone1982}%
  \BibitemOpen
  \bibfield  {author} {\bibinfo {author} {\bibfnamefont {A.}~\bibnamefont
  {Barone}}\ and\ \bibinfo {author} {\bibfnamefont {G.}~\bibnamefont
  {Patern\`{o}}},\ }\href {https://doi.org/10.1002/352760278X} {\emph {\bibinfo
  {title} {{Physics and applications of Josephson effect}}}}\ (\bibinfo
  {publisher} {John Wiley \& Sons},\ \bibinfo {address} {New York},\ \bibinfo
  {year} {1982})\BibitemShut {NoStop}%
\bibitem [{\citenamefont {Lamoreaux}\ \emph {et~al.}(2013)\citenamefont
  {Lamoreaux}, \citenamefont {van Bibber}, \citenamefont {Lehnert},\ and\
  \citenamefont {Carosi}}]{Lamoreaux2013Aug}%
  \BibitemOpen
  \bibfield  {author} {\bibinfo {author} {\bibfnamefont {S.~K.}\ \bibnamefont
  {Lamoreaux}}, \bibinfo {author} {\bibfnamefont {K.~A.}\ \bibnamefont {van
  Bibber}}, \bibinfo {author} {\bibfnamefont {K.~W.}\ \bibnamefont {Lehnert}},\
  and\ \bibinfo {author} {\bibfnamefont {G.}~\bibnamefont {Carosi}},\
  }\bibfield  {title} {\bibinfo {title} {Analysis of single-photon and linear
  amplifier detectors for microwave cavity dark matter axion searches},\ }\href
  {https://doi.org/10.1103/PhysRevD.88.035020} {\bibfield  {journal} {\bibinfo
  {journal} {Phys. Rev. D}\ }\textbf {\bibinfo {volume} {88}},\ \bibinfo
  {pages} {035020} (\bibinfo {year} {2013})}\BibitemShut {NoStop}%
\bibitem [{\citenamefont {Irastorza}\ and\ \citenamefont
  {Redondo}(2018)}]{Irastorza2018Sep}%
  \BibitemOpen
  \bibfield  {author} {\bibinfo {author} {\bibfnamefont {I.~G.}\ \bibnamefont
  {Irastorza}}\ and\ \bibinfo {author} {\bibfnamefont {J.}~\bibnamefont
  {Redondo}},\ }\bibfield  {title} {\bibinfo {title} {New experimental
  approaches in the search for axion-like particles},\ }\href
  {https://doi.org/10.1016/j.ppnp.2018.05.003} {\bibfield  {journal} {\bibinfo
  {journal} {Progress in Particle and Nuclear Physics}\ }\textbf {\bibinfo
  {volume} {102}},\ \bibinfo {pages} {89} (\bibinfo {year} {2018})}\BibitemShut
  {NoStop}%
\bibitem [{\citenamefont {Sikivie}(2021)}]{Sikivie2021Feb}%
  \BibitemOpen
  \bibfield  {author} {\bibinfo {author} {\bibfnamefont {P.}~\bibnamefont
  {Sikivie}},\ }\bibfield  {title} {\bibinfo {title} {{Invisible axion search
  methods}},\ }\href {https://doi.org/10.1103/RevModPhys.93.015004} {\bibfield
  {journal} {\bibinfo  {journal} {Rev. Mod. Phys.}\ }\textbf {\bibinfo {volume}
  {93}},\ \bibinfo {pages} {015004} (\bibinfo {year} {2021})}\BibitemShut
  {NoStop}%
\bibitem [{\citenamefont {Braine}\ \emph {et~al.}(2020)\citenamefont {Braine},
  \citenamefont {Cervantes}, \citenamefont {Crisosto}, \citenamefont {Du},
  \citenamefont {Kimes}, \citenamefont {Rosenberg}, \citenamefont {Rybka},
  \citenamefont {Yang}, \citenamefont {Bowring}, \citenamefont {Chou},
  \citenamefont {Khatiwada}, \citenamefont {Sonnenschein}, \citenamefont
  {Wester}, \citenamefont {Carosi}, \citenamefont {Woollett}, \citenamefont
  {Duffy}, \citenamefont {Bradley}, \citenamefont {Boutan}, \citenamefont
  {Jones}, \citenamefont {LaRoque}, \citenamefont {Oblath}, \citenamefont
  {Taubman}, \citenamefont {Clarke}, \citenamefont {Dove}, \citenamefont
  {Eddins}, \citenamefont {O'Kelley}, \citenamefont {Nawaz}, \citenamefont
  {Siddiqi}, \citenamefont {Stevenson}, \citenamefont {Agrawal}, \citenamefont
  {Dixit}, \citenamefont {Gleason}, \citenamefont {Jois}, \citenamefont
  {Sikivie}, \citenamefont {Solomon}, \citenamefont {Sullivan}, \citenamefont
  {Tanner}, \citenamefont {Lentz}, \citenamefont {Daw}, \citenamefont
  {Buckley}, \citenamefont {Harrington}, \citenamefont {Henriksen},\ and\
  \citenamefont {Murch}}]{PhysRevLett.124.101303}%
  \BibitemOpen
  \bibfield  {author} {\bibinfo {author} {\bibfnamefont {T.}~\bibnamefont
  {Braine}} \emph {et~al.} (\bibinfo {collaboration} {ADMX Collaboration})
  }\href {https://doi.org/10.1103/PhysRevLett.124.101303} {\bibfield  {journal}
  {\bibinfo  {journal} {Phys. Rev. Lett.}\ }\textbf {\bibinfo {volume} {124}},\
  \bibinfo {pages} {101303} (\bibinfo {year} {2020})}\BibitemShut {NoStop}%
\bibitem [{\citenamefont {Crescini}\ \emph {et~al.}(2020)\citenamefont
  {Crescini}, \citenamefont {Alesini}, \citenamefont {Braggio}, \citenamefont
  {Carugno}, \citenamefont {D{'}Agostino}, \citenamefont {Di~Gioacchino},
  \citenamefont {Falferi}, \citenamefont {Gambardella}, \citenamefont {Gatti},
  \citenamefont {Iannone}, \citenamefont {Ligi}, \citenamefont {Lombardi},
  \citenamefont {Ortolan}, \citenamefont {Pengo}, \citenamefont {Ruoso},\ and\
  \citenamefont {Taffarello}}]{Crescini2020May}%
  \BibitemOpen
  \bibfield  {author} {\bibinfo {author} {\bibfnamefont {N.}~\bibnamefont
  {Crescini}}, \bibinfo {author} {\bibfnamefont {D.}~\bibnamefont {Alesini}},
  \bibinfo {author} {\bibfnamefont {C.}~\bibnamefont {Braggio}}, \bibinfo
  {author} {\bibfnamefont {G.}~\bibnamefont {Carugno}}, \bibinfo {author}
  {\bibfnamefont {D.}~\bibnamefont {D{'}Agostino}}, \bibinfo {author}
  {\bibfnamefont {D.}~\bibnamefont {Di~Gioacchino}}, \bibinfo {author}
  {\bibfnamefont {P.}~\bibnamefont {Falferi}}, \bibinfo {author} {\bibfnamefont
  {U.}~\bibnamefont {Gambardella}}, \bibinfo {author} {\bibfnamefont
  {C.}~\bibnamefont {Gatti}}, \bibinfo {author} {\bibfnamefont
  {G.}~\bibnamefont {Iannone}}, \bibinfo {author} {\bibfnamefont
  {C.}~\bibnamefont {Ligi}}, \bibinfo {author} {\bibfnamefont {A.}~\bibnamefont
  {Lombardi}}, \bibinfo {author} {\bibfnamefont {A.}~\bibnamefont {Ortolan}},
  \bibinfo {author} {\bibfnamefont {R.}~\bibnamefont {Pengo}}, \bibinfo
  {author} {\bibfnamefont {G.}~\bibnamefont {Ruoso}},\ and\ \bibinfo {author}
  {\bibfnamefont {L.}~\bibnamefont {Taffarello}},\ }\bibfield  {title}
  {\bibinfo {title} {{Axion Search with a Quantum-Limited Ferromagnetic
  Haloscope}},\ }\href {https://doi.org/10.1103/PhysRevLett.124.171801}
  {\bibfield  {journal} {\bibinfo  {journal} {Phys. Rev. Lett.}\ }\textbf
  {\bibinfo {volume} {124}},\ \bibinfo {pages} {171801} (\bibinfo {year}
  {2020})}\BibitemShut {NoStop}%
\bibitem [{\citenamefont {Kwon}\ \emph {et~al.}(2021)\citenamefont {Kwon},
  \citenamefont {Lee}, \citenamefont {Chung}, \citenamefont {Ahn},
  \citenamefont {Byun}, \citenamefont {Caspers}, \citenamefont {Choi},
  \citenamefont {Choi}, \citenamefont {Chong}, \citenamefont {Jeong},
  \citenamefont {Jeong}, \citenamefont {Kim}, \citenamefont {Kim},
  \citenamefont {Kutlu}, \citenamefont {Lee}, \citenamefont {Lee},
  \citenamefont {Lee}, \citenamefont {Matlashov}, \citenamefont {Oh},
  \citenamefont {Park}, \citenamefont {Uchaikin}, \citenamefont {Youn},\ and\
  \citenamefont {Semertzidis}}]{Kwon2021May}%
  \BibitemOpen
  \bibfield  {author} {\bibinfo {author} {\bibfnamefont {O.}~\bibnamefont
  {Kwon}} \emph {et~al.}, \ }
  \href{https://doi.org/10.1103/PhysRevLett.126.191802} {\bibfield  {journal}
  {\bibinfo  {journal} {Phys. Rev. Lett.}\ }\textbf {\bibinfo {volume} {126}},\
  \bibinfo {pages} {191802} (\bibinfo {year} {2021})}\BibitemShut {NoStop}%
\bibitem [{\citenamefont {Adair}\ \emph {et~al.}(2022)\citenamefont {Adair},
  \citenamefont {Altenm{\"u}ller}, \citenamefont {Anastassopoulos},
  \citenamefont {Arguedas~Cuendis}, \citenamefont {Baier}, \citenamefont
  {Barth}, \citenamefont {Belov}, \citenamefont {Bozicevic}, \citenamefont
  {Br{\"a}uninger}, \citenamefont {Cantatore}, \citenamefont {Caspers},
  \citenamefont {Castel}, \citenamefont {{\c{C}}etin}, \citenamefont {Chung},
  \citenamefont {Choi}, \citenamefont {Choi}, \citenamefont {Dafni},
  \citenamefont {Davenport}, \citenamefont {Dermenev}, \citenamefont {Desch},
  \citenamefont {D{\"o}brich}, \citenamefont {Fischer}, \citenamefont {Funk},
  \citenamefont {Galan}, \citenamefont {Gardikiotis}, \citenamefont {Gninenko},
  \citenamefont {Golm}, \citenamefont {Hasinoff}, \citenamefont {Hoffmann},
  \citenamefont {D{\'i}ez~Ib{\'a}{\~{n}}ez}, \citenamefont {Irastorza},
  \citenamefont {Jakov{\v{c}}i{\'{c}}}, \citenamefont {Kaminski}, \citenamefont
  {Karuza}, \citenamefont {Krieger}, \citenamefont {Kutlu}, \citenamefont
  {Laki{\'{c}}}, \citenamefont {Laurent}, \citenamefont {Lee}, \citenamefont
  {Lee}, \citenamefont {Luz{\'o}n}, \citenamefont {Malbrunot}, \citenamefont
  {Margalejo}, \citenamefont {Maroudas}, \citenamefont {Miceli}, \citenamefont
  {Mirallas}, \citenamefont {Obis}, \citenamefont {{\"O}zbey}, \citenamefont
  {{\"O}zbozduman}, \citenamefont {Pivovaroff}, \citenamefont {Rosu},
  \citenamefont {Ruz}, \citenamefont {Ruiz-Ch{\'o}liz}, \citenamefont
  {Schmidt}, \citenamefont {Schumann}, \citenamefont {Semertzidis},
  \citenamefont {Solanki}, \citenamefont {Stewart}, \citenamefont {Tsagris},
  \citenamefont {Vafeiadis}, \citenamefont {Vogel}, \citenamefont {Vretenar},
  \citenamefont {Youn},\ and\ \citenamefont {Zioutas}}]{Adair2022Oct}%
  \BibitemOpen
  \bibfield  {author} {\bibinfo {author} {\bibfnamefont {C.~M.}\ \bibnamefont
  {Adair}} \emph {et~al.},\ }\href
  {https://doi.org/10.1038/s41467-022-33913-6} {\bibfield  {journal} {\bibinfo
  {journal} {Nature Communications}\ }\textbf {\bibinfo {volume} {13}},\
  \bibinfo {pages} {6180} (\bibinfo {year} {2022})}\BibitemShut {NoStop}%
\bibitem [{\citenamefont {Sushkov}(2023)}]{Sushkov2023May}%
  \BibitemOpen
  \bibfield  {author} {\bibinfo {author} {\bibfnamefont {A.~O.}\ \bibnamefont
  {Sushkov}},\ }\bibfield  {title} {\bibinfo {title} {{Quantum Science and the
  Search for Axion Dark Matter}},\ }\href
  {https://doi.org/10.1103/PRXQuantum.4.020101} {\bibfield  {journal} {\bibinfo
   {journal} {Phys. Rev. X Quantum}\ }\textbf {\bibinfo {volume} {4}},\
  \bibinfo {pages} {020101} (\bibinfo {year} {2023})}\BibitemShut {NoStop}%
\bibitem [{\citenamefont {Graham}\ \emph {et~al.}(2024)\citenamefont {Graham},
  \citenamefont {Ghosh}, \citenamefont {Zhu}, \citenamefont {Bai},
  \citenamefont {Cahn}, \citenamefont {Durcan}, \citenamefont {Jewell},
  \citenamefont {Speller}, \citenamefont {Zacarias}, \citenamefont {Zhou},\
  and\ \citenamefont {Maruyama}}]{Graham2024Feb}%
  \BibitemOpen
  \bibfield  {author} {\bibinfo {author} {\bibfnamefont {E.}~\bibnamefont
  {Graham}}, \bibinfo {author} {\bibfnamefont {S.}~\bibnamefont {Ghosh}},
  \bibinfo {author} {\bibfnamefont {Y.}~\bibnamefont {Zhu}}, \bibinfo {author}
  {\bibfnamefont {X.}~\bibnamefont {Bai}}, \bibinfo {author} {\bibfnamefont
  {S.~B.}\ \bibnamefont {Cahn}}, \bibinfo {author} {\bibfnamefont
  {E.}~\bibnamefont {Durcan}}, \bibinfo {author} {\bibfnamefont {M.~J.}\
  \bibnamefont {Jewell}}, \bibinfo {author} {\bibfnamefont {D.~H.}\
  \bibnamefont {Speller}}, \bibinfo {author} {\bibfnamefont {S.~M.}\
  \bibnamefont {Zacarias}}, \bibinfo {author} {\bibfnamefont {L.~T.}\
  \bibnamefont {Zhou}},\ and\ \bibinfo {author} {\bibfnamefont {R.~H.}\
  \bibnamefont {Maruyama}},\ }\bibfield  {title} {\bibinfo {title}
  {{Rydberg-atom-based single-photon detection for haloscope axion searches}},\
  }\href {https://doi.org/10.1103/PhysRevD.109.032009} {\bibfield  {journal}
  {\bibinfo  {journal} {Phys. Rev. D}\ }\textbf {\bibinfo {volume} {109}},\
  \bibinfo {pages} {032009} (\bibinfo {year} {2024})}\BibitemShut {NoStop}%
\bibitem [{\citenamefont {Buschmann}\ \emph {et~al.}(2022)\citenamefont
  {Buschmann}, \citenamefont {Foster}, \citenamefont {Hook}, \citenamefont
  {Peterson}, \citenamefont {Willcox}, \citenamefont {Zhang},\ and\
  \citenamefont {Safdi}}]{Buschmann2022Feb}%
  \BibitemOpen
  \bibfield  {author} {\bibinfo {author} {\bibfnamefont {M.}~\bibnamefont
  {Buschmann}}, \bibinfo {author} {\bibfnamefont {J.~W.}\ \bibnamefont
  {Foster}}, \bibinfo {author} {\bibfnamefont {A.}~\bibnamefont {Hook}},
  \bibinfo {author} {\bibfnamefont {A.}~\bibnamefont {Peterson}}, \bibinfo
  {author} {\bibfnamefont {D.~E.}\ \bibnamefont {Willcox}}, \bibinfo {author}
  {\bibfnamefont {W.}~\bibnamefont {Zhang}},\ and\ \bibinfo {author}
  {\bibfnamefont {B.~R.}\ \bibnamefont {Safdi}},\ }\bibfield  {title} {\bibinfo
  {title} {{Dark matter from axion strings with adaptive mesh refinement}},\
  }\bibfield  {journal} {\bibinfo  {journal} {Nat. Commun.}\ }\textbf {\bibinfo
  {volume} {13:1049}},\ \href {https://doi.org/10.1038/s41467-022-28669-y}
  (\bibinfo {year} {2022})\BibitemShut {NoStop}%
\bibitem [{\citenamefont {Chen}\ \emph {et~al.}(2011)\citenamefont {Chen},
  \citenamefont {Hover}, \citenamefont {Sendelbach}, \citenamefont {Maurer},
  \citenamefont {Merkel}, \citenamefont {Pritchett}, \citenamefont {Wilhelm},\
  and\ \citenamefont {McDermott}}]{Chen2011Nov}%
  \BibitemOpen
  \bibfield  {author} {\bibinfo {author} {\bibfnamefont {Y.-F.}\ \bibnamefont
  {Chen}}, \bibinfo {author} {\bibfnamefont {D.}~\bibnamefont {Hover}},
  \bibinfo {author} {\bibfnamefont {S.}~\bibnamefont {Sendelbach}}, \bibinfo
  {author} {\bibfnamefont {L.}~\bibnamefont {Maurer}}, \bibinfo {author}
  {\bibfnamefont {S.~T.}\ \bibnamefont {Merkel}}, \bibinfo {author}
  {\bibfnamefont {E.~J.}\ \bibnamefont {Pritchett}}, \bibinfo {author}
  {\bibfnamefont {F.~K.}\ \bibnamefont {Wilhelm}},\ and\ \bibinfo {author}
  {\bibfnamefont {R.}~\bibnamefont {McDermott}},\ }\bibfield  {title}
  {\bibinfo {title} {{Microwave Photon Counter Based on Josephson Junctions}},\ }\href
  {https://doi.org/10.1103/PhysRevLett.107.217401} {\bibfield  {journal}
  {\bibinfo  {journal} {Phys. Rev. Lett.}\ }\textbf {\bibinfo {volume} {107}},\
  \bibinfo {pages} {217401} (\bibinfo {year} {2011})}\BibitemShut {NoStop}%
\bibitem [{\citenamefont {Peropadre}\ \emph {et~al.}(2011)\citenamefont
  {Peropadre}, \citenamefont {Romero}, \citenamefont {Johansson}, \citenamefont
  {Wilson}, \citenamefont {Solano},\ and\ \citenamefont
  {Garc'{\i}a-Ripoll}}]{Peropadre2011Dec}%
  \BibitemOpen
  \bibfield  {author} {\bibinfo {author} {\bibfnamefont {B.}~\bibnamefont
  {Peropadre}}, \bibinfo {author} {\bibfnamefont {G.}~\bibnamefont {Romero}},
  \bibinfo {author} {\bibfnamefont {G.}~\bibnamefont {Johansson}}, \bibinfo
  {author} {\bibfnamefont {C.~M.}\ \bibnamefont {Wilson}}, \bibinfo {author}
  {\bibfnamefont {E.}~\bibnamefont {Solano}},\ and\ \bibinfo {author}
  {\bibfnamefont {J.~J.}\ \bibnamefont {Garc'{\i}a-Ripoll}},\ }\bibfield
  {title} {\bibinfo {title} {{Approaching perfect microwave photodetection in
  circuit QED}},\ }\href {https://doi.org/10.1103/PhysRevA.84.063834}
  {\bibfield  {journal} {\bibinfo  {journal} {Phys. Rev. A}\ }\textbf {\bibinfo
  {volume} {84}},\ \bibinfo {pages} {063834} (\bibinfo {year}
  {2011})}\BibitemShut {NoStop}%
\bibitem [{\citenamefont {Addesso}\ \emph {et~al.}(2012)\citenamefont
  {Addesso}, \citenamefont {Filatrella},\ and\ \citenamefont
  {Pierro}}]{Addesso2012Jan}%
  \BibitemOpen
  \bibfield  {author} {\bibinfo {author} {\bibfnamefont {P.}~\bibnamefont
  {Addesso}}, \bibinfo {author} {\bibfnamefont {G.}~\bibnamefont
  {Filatrella}},\ and\ \bibinfo {author} {\bibfnamefont {V.}~\bibnamefont
  {Pierro}},\ }\bibfield  {title} {\bibinfo {title}
  {{Characterization of escape times of Josephson junctions for signal detection}},\ }\href {https://doi.org/10.1103/PhysRevE.85.016708} {\bibfield
   {journal} {\bibinfo  {journal} {Physical Review E}\ }\textbf {\bibinfo
  {volume} {85}},\ \bibinfo {pages} {016708} (\bibinfo {year}
  {2012})}\BibitemShut {NoStop}%
\bibitem [{\citenamefont {Poudel}\ \emph {et~al.}(2012)\citenamefont {Poudel},
  \citenamefont {McDermott},\ and\ \citenamefont {Vavilov}}]{Poudel2012Nov}%
  \BibitemOpen
  \bibfield  {author} {\bibinfo {author} {\bibfnamefont {A.}~\bibnamefont
  {Poudel}}, \bibinfo {author} {\bibfnamefont {R.}~\bibnamefont {McDermott}},\
  and\ \bibinfo {author} {\bibfnamefont {M.~G.}\ \bibnamefont {Vavilov}},\
  }\bibfield  {title} {\bibinfo {title} {{Quantum
  efficiency of a microwave photon detector based on a current-biased Josephson junction}},\ }\href {https://doi.org/10.1103/PhysRevB.86.174506} {\bibfield
  {journal} {\bibinfo  {journal} {Phys. Rev. B}\ }\textbf {\bibinfo {volume}
  {86}},\ \bibinfo {pages} {174506} (\bibinfo {year} {2012})}\BibitemShut
  {NoStop}%
\bibitem [{\citenamefont {Kuzmin}\ \emph {et~al.}(2018)\citenamefont {Kuzmin},
  \citenamefont {Sobolev}, \citenamefont {Gatti}, \citenamefont
  {Di~Gioacchino}, \citenamefont {Crescini}, \citenamefont {Gordeeva},\ and\
  \citenamefont {Il'ichev}}]{Kuzmin2018Jun}%
  \BibitemOpen
  \bibfield  {author} {\bibinfo {author} {\bibfnamefont {L.~S.}\ \bibnamefont
  {Kuzmin}}, \bibinfo {author} {\bibfnamefont {A.~S.}\ \bibnamefont {Sobolev}},
  \bibinfo {author} {\bibfnamefont {C.}~\bibnamefont {Gatti}}, \bibinfo
  {author} {\bibfnamefont {D.}~\bibnamefont {Di~Gioacchino}}, \bibinfo {author}
  {\bibfnamefont {N.}~\bibnamefont {Crescini}}, \bibinfo {author}
  {\bibfnamefont {A.}~\bibnamefont {Gordeeva}},\ and\ \bibinfo {author}
  {\bibfnamefont {E.}~\bibnamefont {Il'ichev}},\ }\bibfield  {title} {\bibinfo
  {title} {{Single Photon Counter Based on a Josephson Junction at 14 GHz for
  Searching Galactic Axions}},\ }\href
  {https://doi.org/10.1109/TASC.2018.2850019} {\bibfield  {journal} {\bibinfo
  {journal} {IEEE Trans. Appl. Supercond.}\ }\textbf {\bibinfo {volume} {28}},\
  \bibinfo {pages} {2400505} (\bibinfo {year} {2018})}\BibitemShut {NoStop}%
\bibitem [{\citenamefont {Guarcello}\ \emph
  {et~al.}(2019{\natexlab{a}})\citenamefont {Guarcello}, \citenamefont
  {Valenti}, \citenamefont {Spagnolo}, \citenamefont {Pierro},\ and\
  \citenamefont {Filatrella}}]{Guarcello2019Apr}%
  \BibitemOpen
  \bibfield  {author} {\bibinfo {author} {\bibfnamefont {C.}~\bibnamefont
  {Guarcello}}, \bibinfo {author} {\bibfnamefont {D.}~\bibnamefont {Valenti}},
  \bibinfo {author} {\bibfnamefont {B.}~\bibnamefont {Spagnolo}}, \bibinfo
  {author} {\bibfnamefont {V.}~\bibnamefont {Pierro}},\ and\ \bibinfo {author}
  {\bibfnamefont {G.}~\bibnamefont {Filatrella}},\ }\bibfield  {title}
  {\bibinfo {title} {{Josephson-based Threshold Detector for L\'{e}vy-Distributed Current Fluctuations}},\ }\href
  {https://doi.org/10.1103/PhysRevApplied.11.044078} {\bibfield  {journal}
  {\bibinfo  {journal} {Physical Review Applied}\ }\textbf {\bibinfo {volume}
  {11}},\ \bibinfo {pages} {044078} (\bibinfo {year}
  {2019}{\natexlab{a}})}\BibitemShut {NoStop}%
\bibitem [{\citenamefont {Guarcello}\ \emph
  {et~al.}(2019{\natexlab{b}})\citenamefont {Guarcello}, \citenamefont
  {Braggio}, \citenamefont {Solinas}, \citenamefont {Pepe},\ and\ \citenamefont
  {Giazotto}}]{Guarcello2019May}%
  \BibitemOpen
  \bibfield  {author} {\bibinfo {author} {\bibfnamefont {C.}~\bibnamefont
  {Guarcello}}, \bibinfo {author} {\bibfnamefont {A.}~\bibnamefont {Braggio}},
  \bibinfo {author} {\bibfnamefont {P.}~\bibnamefont {Solinas}}, \bibinfo
  {author} {\bibfnamefont {G.~P.}\ \bibnamefont {Pepe}},\ and\ \bibinfo
  {author} {\bibfnamefont {F.}~\bibnamefont {Giazotto}},\ }\bibfield  {title}
  {\bibinfo {title} {{Josephson-Threshold
  Calorimeter}},\ }\href {https://doi.org/10.1103/PhysRevApplied.11.054074}
  {\bibfield  {journal} {\bibinfo  {journal} {Physical Review Applied}\
  }\textbf {\bibinfo {volume} {11}},\ \bibinfo {pages} {054074} (\bibinfo
  {year} {2019}{\natexlab{b}})}\BibitemShut {NoStop}%
\bibitem [{\citenamefont {Yablokov}\ \emph {et~al.}(2020)\citenamefont
  {Yablokov}, \citenamefont {Mylnikov}, \citenamefont {Pankratov},
  \citenamefont {Pankratova},\ and\ \citenamefont
  {Gordeeva}}]{Yablokov2020Jul}%
  \BibitemOpen
  \bibfield  {author} {\bibinfo {author} {\bibfnamefont {A.~A.}\ \bibnamefont
  {Yablokov}}, \bibinfo {author} {\bibfnamefont {V.~M.}\ \bibnamefont
  {Mylnikov}}, \bibinfo {author} {\bibfnamefont {A.~L.}\ \bibnamefont
  {Pankratov}}, \bibinfo {author} {\bibfnamefont {E.~V.}\ \bibnamefont
  {Pankratova}},\ and\ \bibinfo {author} {\bibfnamefont {A.~V.}\ \bibnamefont
  {Gordeeva}},\ }\bibfield  {title} {\bibinfo {title}
  {{Suppression of switching errors in weakly damped Josephson junctions}},\
  }\href {https://doi.org/10.1016/j.chaos.2020.109817} {\bibfield  {journal}
  {\bibinfo  {journal} {Chaos, Solitons {\&} Fractals}\ }\textbf {\bibinfo
  {volume} {136}},\ \bibinfo {pages} {109817} (\bibinfo {year}
  {2020})}\BibitemShut {NoStop}%
\bibitem [{\citenamefont {Piedjou~Komnang}\ \emph {et~al.}(2021)\citenamefont
  {Piedjou~Komnang}, \citenamefont {Guarcello}, \citenamefont {Barone},
  \citenamefont {Gatti}, \citenamefont {Pagano}, \citenamefont {Pierro},
  \citenamefont {Rettaroli},\ and\ \citenamefont
  {Filatrella}}]{PiedjouKomnang2021Jan}%
  \BibitemOpen
  \bibfield  {author} {\bibinfo {author} {\bibfnamefont {A.~S.}\ \bibnamefont
  {Piedjou~Komnang}}, \bibinfo {author} {\bibfnamefont {C.}~\bibnamefont
  {Guarcello}}, \bibinfo {author} {\bibfnamefont {C.}~\bibnamefont {Barone}},
  \bibinfo {author} {\bibfnamefont {C.}~\bibnamefont {Gatti}}, \bibinfo
  {author} {\bibfnamefont {S.}~\bibnamefont {Pagano}}, \bibinfo {author}
  {\bibfnamefont {V.}~\bibnamefont {Pierro}}, \bibinfo {author} {\bibfnamefont
  {A.}~\bibnamefont {Rettaroli}},\ and\ \bibinfo {author} {\bibfnamefont
  {G.}~\bibnamefont {Filatrella}},\ }\bibfield  {title} {{\bibinfo {title} {{Analysis of Josephson junctions switching time distributions for the detection of single microwave photons}}} }\href
  {https://doi.org/10.1016/j.chaos.2020.110496} {\bibfield  {journal} {\bibinfo
   {journal} {Chaos, Solitons {\&} Fractals}\ }\textbf {\bibinfo {volume}
  {142}},\ \bibinfo {pages} {110496} (\bibinfo {year} {2021})}\BibitemShut
  {NoStop}%
\bibitem [{\citenamefont {Yablokov}\ \emph {et~al.}(2021)\citenamefont
  {Yablokov}, \citenamefont {Glushkov}, \citenamefont {Pankratov},
  \citenamefont {Gordeeva}, \citenamefont {Kuzmin},\ and\ \citenamefont
  {Il{'}ichev}}]{Yablokov2021Jul}%
  \BibitemOpen
  \bibfield  {author} {\bibinfo {author} {\bibfnamefont {A.~A.}\ \bibnamefont
  {Yablokov}}, \bibinfo {author} {\bibfnamefont {E.~I.}\ \bibnamefont
  {Glushkov}}, \bibinfo {author} {\bibfnamefont {A.~L.}\ \bibnamefont
  {Pankratov}}, \bibinfo {author} {\bibfnamefont {A.~V.}\ \bibnamefont
  {Gordeeva}}, \bibinfo {author} {\bibfnamefont {L.~S.}\ \bibnamefont
  {Kuzmin}},\ and\ \bibinfo {author} {\bibfnamefont {E.~V.}\ \bibnamefont
  {Il{'}ichev}},\ }\bibfield  {title} {\bibinfo
  {title} {{Resonant response drives sensitivity of Josephson escape detector}},\ }\href {https://doi.org/10.1016/j.chaos.2021.111058} {\bibfield
   {journal} {\bibinfo  {journal} {Chaos, Solitons {\&} Fractals}\ }\textbf
  {\bibinfo {volume} {148}},\ \bibinfo {pages} {111058} (\bibinfo {year}
  {2021})}\BibitemShut {NoStop}%
\bibitem [{\citenamefont {Golubev}\ \emph {et~al.}(2021)\citenamefont
  {Golubev}, \citenamefont {Il{'}ichev},\ and\ \citenamefont
  {Kuzmin}}]{Golubev2021Jul}%
  \BibitemOpen
  \bibfield  {author} {\bibinfo {author} {\bibfnamefont {D.~S.}\ \bibnamefont
  {Golubev}}, \bibinfo {author} {\bibfnamefont {E.~V.}\ \bibnamefont
  {Il{'}ichev}},\ and\ \bibinfo {author} {\bibfnamefont {L.~S.}\ \bibnamefont
  {Kuzmin}},\ }\bibfield  {title} {\bibinfo {title}
  {{Single-Photon Detection with a Josephson Junction Coupled to a Resonator}},\ }\href {https://doi.org/10.1103/PhysRevApplied.16.014025}
  {\bibfield  {journal} {\bibinfo  {journal} {Phys. Rev. Appl.}\ }\textbf
  {\bibinfo {volume} {16}},\ \bibinfo {pages} {014025} (\bibinfo {year}
  {2021})}\BibitemShut {NoStop}%
\bibitem [{\citenamefont {Ladeynov}\ \emph {et~al.}(2023)\citenamefont
  {Ladeynov}, \citenamefont {Egorov},\ and\ \citenamefont
  {Pankratov}}]{Ladeynov2023Jun}%
  \BibitemOpen
  \bibfield  {author} {\bibinfo {author} {\bibfnamefont {D.~A.}\ \bibnamefont
  {Ladeynov}}, \bibinfo {author} {\bibfnamefont {D.~G.}\ \bibnamefont
  {Egorov}},\ and\ \bibinfo {author} {\bibfnamefont {A.~L.}\ \bibnamefont
  {Pankratov}},\ }\bibfield  {title} {\bibinfo {title} {{Stochastic versus
  dynamic resonant activation to enhance threshold detector sensitivity}},\
  }\href {https://doi.org/10.1016/j.chaos.2023.113506} {\bibfield  {journal}
  {\bibinfo  {journal} {Chaos, Solitons \& Fractals}\ }\textbf {\bibinfo
  {volume} {171}},\ \bibinfo {pages} {113506} (\bibinfo {year}
  {2023})}\BibitemShut {NoStop}%
\bibitem [{\citenamefont {Stanisavljevi{'{c}}}\ \emph
  {et~al.}(2024)\citenamefont {Stanisavljevi{'{c}}}, \citenamefont {Philippe},
  \citenamefont {Gabelli}, \citenamefont {Aprili}, \citenamefont
  {Est{`{e}}ve},\ and\ \citenamefont {Basset}}]{Stanisavljevic2024Aug}%
  \BibitemOpen
  \bibfield  {author} {\bibinfo {author} {\bibfnamefont {O.}~\bibnamefont
  {Stanisavljevi{'{c}}}}, \bibinfo {author} {\bibfnamefont {J.-C.}\
  \bibnamefont {Philippe}}, \bibinfo {author} {\bibfnamefont {J.}~\bibnamefont
  {Gabelli}}, \bibinfo {author} {\bibfnamefont {M.}~\bibnamefont {Aprili}},
  \bibinfo {author} {\bibfnamefont {J.}~\bibnamefont {Est{`{e}}ve}},\ and\
  \bibinfo {author} {\bibfnamefont {J.}~\bibnamefont {Basset}},\ }\bibfield
  {title} {\bibinfo {title} {{Efficient Microwave Photon-to-Electron Conversion
  in a High-Impedance Quantum Circuit}},\ }\href
  {https://doi.org/10.1103/PhysRevLett.133.076302} {\bibfield  {journal}
  {\bibinfo  {journal} {Phys. Rev. Lett.}\ }\textbf {\bibinfo {volume} {133}},\
  \bibinfo {pages} {076302} (\bibinfo {year} {2024})}\BibitemShut {NoStop}%
\bibitem [{\citenamefont {Pankratov}\ \emph
  {et~al.}(2022{\natexlab{a}})\citenamefont {Pankratov}, \citenamefont {Revin},
  \citenamefont {Gordeeva}, \citenamefont {Yablokov}, \citenamefont {Kuzmin},\
  and\ \citenamefont {Ilichev}}]{Pankratov2022May}%
  \BibitemOpen
  \bibfield  {author} {\bibinfo {author} {\bibfnamefont {A.~L.}\ \bibnamefont
  {Pankratov}}, \bibinfo {author} {\bibfnamefont {L.~S.}\ \bibnamefont
  {Revin}}, \bibinfo {author} {\bibfnamefont {A.~V.}\ \bibnamefont {Gordeeva}},
  \bibinfo {author} {\bibfnamefont {A.~A.}\ \bibnamefont {Yablokov}}, \bibinfo
  {author} {\bibfnamefont {L.~S.}\ \bibnamefont {Kuzmin}},\ and\ \bibinfo
  {author} {\bibfnamefont {E.~V.}\ \bibnamefont {Ilichev}},\ }\bibfield
  {title} {\bibinfo {title} {{Towards a microwave single-photon counter for searching axions}},\ }\href
  {https://doi.org/10.1038/s41534-022-00569-5} {\bibfield  {journal} {\bibinfo
  {journal} {npj Quantum Information}\ }\textbf {\bibinfo {volume} {8}},\
  \bibinfo {pages} {1} (\bibinfo {year} {2022}{\natexlab{a}})}\BibitemShut
  {NoStop}%
\bibitem [{\citenamefont {Martinis}\ and\ \citenamefont
  {Kautz}(1989)}]{Martinis1989Oct}%
  \BibitemOpen
  \bibfield  {author} {\bibinfo {author} {\bibfnamefont {J.~M.}\ \bibnamefont
  {Martinis}}\ and\ \bibinfo {author} {\bibfnamefont {R.~L.}\ \bibnamefont
  {Kautz}},\ }\bibfield  {title} {\bibinfo {title} {{Classical phase diffusion
  in small hysteretic Josephson junctions}},\ }\href
  {https://doi.org/10.1103/PhysRevLett.63.1507} {\bibfield  {journal} {\bibinfo
   {journal} {Phys. Rev. Lett.}\ }\textbf {\bibinfo {volume} {63}},\ \bibinfo
  {pages} {1507} (\bibinfo {year} {1989})}\BibitemShut {NoStop}%
\bibitem [{\citenamefont {Kautz}\ and\ \citenamefont
  {Martinis}(1990)}]{Kautz1990Dec}%
  \BibitemOpen
  \bibfield  {author} {\bibinfo {author} {\bibfnamefont {R.~L.}\ \bibnamefont
  {Kautz}}\ and\ \bibinfo {author} {\bibfnamefont {J.~M.}\ \bibnamefont
  {Martinis}},\ }\bibfield  {title} {\bibinfo {title} {{Noise-affected I-V
  curves in small hysteretic Josephson junctions}},\ }\href
  {https://doi.org/10.1103/PhysRevB.42.9903} {\bibfield  {journal} {\bibinfo
  {journal} {Phys. Rev. B}\ }\textbf {\bibinfo {volume} {42}},\ \bibinfo
  {pages} {9903} (\bibinfo {year} {1990})}\BibitemShut {NoStop}%
\bibitem [{\citenamefont {Koval}\ \emph {et~al.}(2004)\citenamefont {Koval},
  \citenamefont {Fistul},\ and\ \citenamefont {Ustinov}}]{Koval2004Aug}%
  \BibitemOpen
  \bibfield  {author} {\bibinfo {author} {\bibfnamefont {Y.}~\bibnamefont
  {Koval}}, \bibinfo {author} {\bibfnamefont {M.~V.}\ \bibnamefont {Fistul}},\
  and\ \bibinfo {author} {\bibfnamefont {A.~V.}\ \bibnamefont {Ustinov}},\
  }\bibfield  {title} {\bibinfo {title} {{Enhancement of Josephson Phase Diffusion by Microwaves}},\ }\href
  {https://doi.org/10.1103/PhysRevLett.93.087004} {\bibfield  {journal}
  {\bibinfo  {journal} {Phys. Rev. Lett.}\ }\textbf {\bibinfo {volume} {93}},\
  \bibinfo {pages} {087004} (\bibinfo {year} {2004})}\BibitemShut {NoStop}%
\bibitem [{\citenamefont {Kivioja}\ \emph {et~al.}(2005)\citenamefont
  {Kivioja}, \citenamefont {Nieminen}, \citenamefont {Claudon}, \citenamefont
  {Buisson}, \citenamefont {Hekking},\ and\ \citenamefont
  {Pekola}}]{Kivioja2005Jun}%
  \BibitemOpen
  \bibfield  {author} {\bibinfo {author} {\bibfnamefont {J.~M.}\ \bibnamefont
  {Kivioja}}, \bibinfo {author} {\bibfnamefont {T.~E.}\ \bibnamefont
  {Nieminen}}, \bibinfo {author} {\bibfnamefont {J.}~\bibnamefont {Claudon}},
  \bibinfo {author} {\bibfnamefont {O.}~\bibnamefont {Buisson}}, \bibinfo
  {author} {\bibfnamefont {F.~W.~J.}\ \bibnamefont {Hekking}},\ and\ \bibinfo
  {author} {\bibfnamefont {J.~P.}\ \bibnamefont {Pekola}},\ }\bibfield  {title}
  {\bibinfo {title} {{Observation of Transition from Escape Dynamics to
  Underdamped Phase Diffusion in a Josephson Junction}},\ }\href
  {https://doi.org/10.1103/PhysRevLett.94.247002} {\bibfield  {journal}
  {\bibinfo  {journal} {Phys. Rev. Lett.}\ }\textbf {\bibinfo {volume} {94}},\
  \bibinfo {pages} {247002} (\bibinfo {year} {2005})}\BibitemShut {NoStop}%
\bibitem [{\citenamefont {M\"{a}nnik}\ \emph {et~al.}(2005)\citenamefont
  {M\"{a}nnik}, \citenamefont {Li}, \citenamefont {Qiu}, \citenamefont {Chen},
  \citenamefont {Patel}, \citenamefont {Han},\ and\ \citenamefont
  {Lukens}}]{Mannik2005Jun}%
  \BibitemOpen
  \bibfield  {author} {\bibinfo {author} {\bibfnamefont {J.}~\bibnamefont
  {M\"{a}nnik}}, \bibinfo {author} {\bibfnamefont {S.}~\bibnamefont {Li}},
  \bibinfo {author} {\bibfnamefont {W.}~\bibnamefont {Qiu}}, \bibinfo {author}
  {\bibfnamefont {W.}~\bibnamefont {Chen}}, \bibinfo {author} {\bibfnamefont
  {V.}~\bibnamefont {Patel}}, \bibinfo {author} {\bibfnamefont
  {S.}~\bibnamefont {Han}},\ and\ \bibinfo {author} {\bibfnamefont {J.~E.}\
  \bibnamefont {Lukens}},\ }\bibfield  {title} {\bibinfo {title} {{Crossover
  from Kramers to phase-diffusion switching in moderately damped Josephson
  junctions}},\ }\href {https://doi.org/10.1103/PhysRevB.71.220509} {\bibfield
  {journal} {\bibinfo  {journal} {Phys. Rev. B}\ }\textbf {\bibinfo {volume}
  {71}},\ \bibinfo {pages} {220509} (\bibinfo {year} {2005})}\BibitemShut
  {NoStop}%
\bibitem [{\citenamefont {Krasnov}\ \emph {et~al.}(2005)\citenamefont
  {Krasnov}, \citenamefont {Bauch}, \citenamefont {Intiso}, \citenamefont
  {H\"{u}rfeld}, \citenamefont {Akazaki}, \citenamefont {Takayanagi},\ and\
  \citenamefont {Delsing}}]{Krasnov2005Oct}%
  \BibitemOpen
  \bibfield  {author} {\bibinfo {author} {\bibfnamefont {V.~M.}\ \bibnamefont
  {Krasnov}}, \bibinfo {author} {\bibfnamefont {T.}~\bibnamefont {Bauch}},
  \bibinfo {author} {\bibfnamefont {S.}~\bibnamefont {Intiso}}, \bibinfo
  {author} {\bibfnamefont {E.}~\bibnamefont {H\"{u}rfeld}}, \bibinfo {author}
  {\bibfnamefont {T.}~\bibnamefont {Akazaki}}, \bibinfo {author} {\bibfnamefont
  {H.}~\bibnamefont {Takayanagi}},\ and\ \bibinfo {author} {\bibfnamefont
  {P.}~\bibnamefont {Delsing}},\ }\bibfield  {title} {\bibinfo {title}
  {{Collapse of Thermal Activation in Moderately Damped Josephson Junctions}},\
  }\href {https://doi.org/10.1103/PhysRevLett.95.157002} {\bibfield  {journal}
  {\bibinfo  {journal} {Phys. Rev. Lett.}\ }\textbf {\bibinfo {volume} {95}},\
  \bibinfo {pages} {157002} (\bibinfo {year} {2005})}\BibitemShut {NoStop}%
\bibitem [{\citenamefont {Krasnov}\ \emph {et~al.}(2007)\citenamefont
  {Krasnov}, \citenamefont {Golod}, \citenamefont {Bauch},\ and\ \citenamefont
  {Delsing}}]{Krasnov2007Dec}%
  \BibitemOpen
  \bibfield  {author} {\bibinfo {author} {\bibfnamefont {V.~M.}\ \bibnamefont
  {Krasnov}}, \bibinfo {author} {\bibfnamefont {T.}~\bibnamefont {Golod}},
  \bibinfo {author} {\bibfnamefont {T.}~\bibnamefont {Bauch}},\ and\ \bibinfo
  {author} {\bibfnamefont {P.}~\bibnamefont {Delsing}},\ }\bibfield  {title}
  {\bibinfo {title} {{Anticorrelation between temperature and fluctuations of
  the switching current in moderately damped Josephson junctions}},\ }\href
  {https://doi.org/10.1103/PhysRevB.76.224517} {\bibfield  {journal} {\bibinfo
  {journal} {Phys. Rev. B}\ }\textbf {\bibinfo {volume} {76}},\ \bibinfo
  {pages} {224517} (\bibinfo {year} {2007})}\BibitemShut {NoStop}%
\bibitem [{\citenamefont {Fenton}\ and\ \citenamefont
  {Warburton}(2008)}]{Fenton2008Aug}%
  \BibitemOpen
  \bibfield  {author} {\bibinfo {author} {\bibfnamefont {J.~C.}\ \bibnamefont
  {Fenton}}\ and\ \bibinfo {author} {\bibfnamefont {P.~A.}\ \bibnamefont
  {Warburton}},\ }\bibfield  {title} {\bibinfo {title} {{Monte Carlo
  simulations of thermal fluctuations in moderately damped Josephson junctions:
  Multiple escape and retrapping, switching- and return-current distributions,
  and hysteresis}},\ }\href {https://doi.org/10.1103/PhysRevB.78.054526}
  {\bibfield  {journal} {\bibinfo  {journal} {Phys. Rev. B}\ }\textbf {\bibinfo
  {volume} {78}},\ \bibinfo {pages} {054526} (\bibinfo {year}
  {2008})}\BibitemShut {NoStop}%
\bibitem [{\citenamefont {Yoon}\ \emph {et~al.}(2011)\citenamefont {Yoon},
  \citenamefont {Gasparinetti}, \citenamefont {M\"{o}tt\"{o}nen},\ and\
  \citenamefont {Pekola}}]{Yoon2011May}%
  \BibitemOpen
  \bibfield  {author} {\bibinfo {author} {\bibfnamefont {Y.}~\bibnamefont
  {Yoon}}, \bibinfo {author} {\bibfnamefont {S.}~\bibnamefont {Gasparinetti}},
  \bibinfo {author} {\bibfnamefont {M.}~\bibnamefont {M\"{o}tt\"{o}nen}},\ and\
  \bibinfo {author} {\bibfnamefont {J.~P.}\ \bibnamefont {Pekola}},\ }\bibfield
   {title} {\bibinfo {title} {{Capacitively Enhanced Thermal Escape in
  Underdamped Josephson Junctions}},\ }\href
  {https://doi.org/10.1007/s10909-011-0344-2} {\bibfield  {journal} {\bibinfo
  {journal} {J. Low Temp. Phys.}\ }\textbf {\bibinfo {volume} {163}},\ \bibinfo
  {pages} {164} (\bibinfo {year} {2011})}\BibitemShut {NoStop}%
\bibitem [{\citenamefont {Yu}\ \emph {et~al.}(2011)\citenamefont {Yu},
  \citenamefont {Zhu}, \citenamefont {Peng}, \citenamefont {Tian},
  \citenamefont {Cui}, \citenamefont {Chen}, \citenamefont {Zheng},
  \citenamefont {Jing}, \citenamefont {Lu}, \citenamefont {Zhao},\ and\
  \citenamefont {Han}}]{Yu2011Aug}%
  \BibitemOpen
  \bibfield  {author} {\bibinfo {author} {\bibfnamefont {H.~F.}\ \bibnamefont
  {Yu}}, \bibinfo {author} {\bibfnamefont {X.~B.}\ \bibnamefont {Zhu}},
  \bibinfo {author} {\bibfnamefont {Z.~H.}\ \bibnamefont {Peng}}, \bibinfo
  {author} {\bibfnamefont {Y.}~\bibnamefont {Tian}}, \bibinfo {author}
  {\bibfnamefont {D.~J.}\ \bibnamefont {Cui}}, \bibinfo {author} {\bibfnamefont
  {G.~H.}\ \bibnamefont {Chen}}, \bibinfo {author} {\bibfnamefont {D.~N.}\
  \bibnamefont {Zheng}}, \bibinfo {author} {\bibfnamefont {X.~N.}\ \bibnamefont
  {Jing}}, \bibinfo {author} {\bibfnamefont {L.}~\bibnamefont {Lu}}, \bibinfo
  {author} {\bibfnamefont {S.~P.}\ \bibnamefont {Zhao}},\ and\ \bibinfo
  {author} {\bibfnamefont {S.}~\bibnamefont {Han}},\ }\bibfield  {title}
  {\bibinfo {title} {{Quantum Phase Diffusion in a Small Underdamped Josephson
  Junction}},\ }\href {https://doi.org/10.1103/PhysRevLett.107.067004}
  {\bibfield  {journal} {\bibinfo  {journal} {Phys. Rev. Lett.}\ }\textbf
  {\bibinfo {volume} {107}},\ \bibinfo {pages} {067004} (\bibinfo {year}
  {2011})}\BibitemShut {NoStop}%
\bibitem [{\citenamefont {Longobardi}\ \emph {et~al.}(2011)\citenamefont
  {Longobardi}, \citenamefont {Massarotti}, \citenamefont {Rotoli},
  \citenamefont {Stornaiuolo}, \citenamefont {Papari}, \citenamefont
  {Kawakami}, \citenamefont {Pepe}, \citenamefont {Barone},\ and\ \citenamefont
  {Tafuri}}]{Longobardi2011Nov}%
  \BibitemOpen
  \bibfield  {author} {\bibinfo {author} {\bibfnamefont {L.}~\bibnamefont
  {Longobardi}}, \bibinfo {author} {\bibfnamefont {D.}~\bibnamefont
  {Massarotti}}, \bibinfo {author} {\bibfnamefont {G.}~\bibnamefont {Rotoli}},
  \bibinfo {author} {\bibfnamefont {D.}~\bibnamefont {Stornaiuolo}}, \bibinfo
  {author} {\bibfnamefont {G.}~\bibnamefont {Papari}}, \bibinfo {author}
  {\bibfnamefont {A.}~\bibnamefont {Kawakami}}, \bibinfo {author}
  {\bibfnamefont {G.~P.}\ \bibnamefont {Pepe}}, \bibinfo {author}
  {\bibfnamefont {A.}~\bibnamefont {Barone}},\ and\ \bibinfo {author}
  {\bibfnamefont {F.}~\bibnamefont {Tafuri}},\ }\bibfield  {title} {\bibinfo
  {title} {{Thermal hopping and retrapping of a Brownian particle in the tilted
  periodic potential of a NbN/MgO/NbN Josephson junction}},\ }\href
  {https://doi.org/10.1103/PhysRevB.84.184504} {\bibfield  {journal} {\bibinfo
  {journal} {Phys. Rev. B}\ }\textbf {\bibinfo {volume} {84}},\ \bibinfo
  {pages} {184504} (\bibinfo {year} {2011})}\BibitemShut {NoStop}%
\bibitem [{\citenamefont {Longobardi}\ \emph {et~al.}(2012)\citenamefont
  {Longobardi}, \citenamefont {Massarotti}, \citenamefont {Stornaiuolo},
  \citenamefont {Galletti}, \citenamefont {Rotoli}, \citenamefont {Lombardi},\
  and\ \citenamefont {Tafuri}}]{Longobardi2012Aug}%
  \BibitemOpen
  \bibfield  {author} {\bibinfo {author} {\bibfnamefont {L.}~\bibnamefont
  {Longobardi}}, \bibinfo {author} {\bibfnamefont {D.}~\bibnamefont
  {Massarotti}}, \bibinfo {author} {\bibfnamefont {D.}~\bibnamefont
  {Stornaiuolo}}, \bibinfo {author} {\bibfnamefont {L.}~\bibnamefont
  {Galletti}}, \bibinfo {author} {\bibfnamefont {G.}~\bibnamefont {Rotoli}},
  \bibinfo {author} {\bibfnamefont {F.}~\bibnamefont {Lombardi}},\ and\
  \bibinfo {author} {\bibfnamefont {F.}~\bibnamefont {Tafuri}},\ }\bibfield
  {title} {\bibinfo {title} {{Direct Transition from Quantum Escape to a Phase
  Diffusion Regime in YBaCuO Biepitaxial Josephson Junctions}},\ }\href
  {https://doi.org/10.1103/PhysRevLett.109.050601} {\bibfield  {journal}
  {\bibinfo  {journal} {Phys. Rev. Lett.}\ }\textbf {\bibinfo {volume} {109}},\
  \bibinfo {pages} {050601} (\bibinfo {year} {2012})}\BibitemShut {NoStop}%
\bibitem [{\citenamefont {Yu}\ \emph {et~al.}(2012)\citenamefont {Yu},
  \citenamefont {Zhu}, \citenamefont {Deng}, \citenamefont {Xue}, \citenamefont
  {Tian}, \citenamefont {Ren}, \citenamefont {Chen}, \citenamefont {Zheng},
  \citenamefont {Jing}, \citenamefont {Lu}, \citenamefont {Zhao},\ and\
  \citenamefont {Han}}]{Yu2012Dec}%
  \BibitemOpen
  \bibfield  {author} {\bibinfo {author} {\bibfnamefont {H.~F.}\ \bibnamefont
  {Yu}}, \bibinfo {author} {\bibfnamefont {X.~B.}\ \bibnamefont {Zhu}},
  \bibinfo {author} {\bibfnamefont {H.}~\bibnamefont {Deng}}, \bibinfo {author}
  {\bibfnamefont {G.~M.}\ \bibnamefont {Xue}}, \bibinfo {author} {\bibfnamefont
  {Y.}~\bibnamefont {Tian}}, \bibinfo {author} {\bibfnamefont {Y.~F.}\
  \bibnamefont {Ren}}, \bibinfo {author} {\bibfnamefont {G.~H.}\ \bibnamefont
  {Chen}}, \bibinfo {author} {\bibfnamefont {D.~N.}\ \bibnamefont {Zheng}},
  \bibinfo {author} {\bibfnamefont {X.~N.}\ \bibnamefont {Jing}}, \bibinfo
  {author} {\bibfnamefont {L.}~\bibnamefont {Lu}}, \bibinfo {author}
  {\bibfnamefont {S.~P.}\ \bibnamefont {Zhao}},\ and\ \bibinfo {author}
  {\bibfnamefont {S.}~\bibnamefont {Han}},\ }\bibfield  {title} {\bibinfo
  {title} {{A two-step transition description of underdamped phase
  diffusion}},\ }\href {https://doi.org/10.1088/1742-6596/400/4/042079}
  {\bibfield  {journal} {\bibinfo  {journal} {J. Phys. Conf. Ser.}\ }\textbf
  {\bibinfo {volume} {400}},\ \bibinfo {pages} {042079} (\bibinfo {year}
  {2012})}\BibitemShut {NoStop}%
\bibitem [{\citenamefont {Revin}\ \emph {et~al.}(2020)\citenamefont {Revin},
  \citenamefont {Pankratov}, \citenamefont {Gordeeva}, \citenamefont
  {Yablokov}, \citenamefont {Rakut}, \citenamefont {Zbrozhek},\ and\
  \citenamefont {Kuzmin}}]{Revin2020Jun}%
  \BibitemOpen
  \bibfield  {author} {\bibinfo {author} {\bibfnamefont {L.~S.}\ \bibnamefont
  {Revin}}, \bibinfo {author} {\bibfnamefont {A.~L.}\ \bibnamefont
  {Pankratov}}, \bibinfo {author} {\bibfnamefont {A.~V.}\ \bibnamefont
  {Gordeeva}}, \bibinfo {author} {\bibfnamefont {A.~A.}\ \bibnamefont
  {Yablokov}}, \bibinfo {author} {\bibfnamefont {I.~V.}\ \bibnamefont {Rakut}},
  \bibinfo {author} {\bibfnamefont {V.~O.}\ \bibnamefont {Zbrozhek}},\ and\
  \bibinfo {author} {\bibfnamefont {L.~S.}\ \bibnamefont {Kuzmin}},\ }\bibfield
   {title} {\bibinfo {title} {Microwave photon detection by an {Al} {Josephson}
  junction},\ }\href {https://doi.org/10.3762/bjnano.11.80} {\bibfield
  {journal} {\bibinfo  {journal} {Beilstein Journal of Nanotechnology}\
  }\textbf {\bibinfo {volume} {11}},\ \bibinfo {pages} {960} (\bibinfo {year}
  {2020})}\BibitemShut {NoStop}%
\bibitem [{\citenamefont {Pankratov}\ \emph {et~al.}(2024)\citenamefont
  {Pankratov}, \citenamefont {Ladeynov}, \citenamefont {Revin}, \citenamefont
  {Gordeeva},\ and\ \citenamefont {Ilichev}}]{Pankratov2024May}%
  \BibitemOpen
  \bibfield  {author} {\bibinfo {author} {\bibfnamefont {A.~L.}\ \bibnamefont
  {Pankratov}}, \bibinfo {author} {\bibfnamefont {D.~A.}\ \bibnamefont
  {Ladeynov}}, \bibinfo {author} {\bibfnamefont {L.~S.}\ \bibnamefont {Revin}},
  \bibinfo {author} {\bibfnamefont {A.~V.}\ \bibnamefont {Gordeeva}},\ and\
  \bibinfo {author} {\bibfnamefont {E.~V.}\ \bibnamefont {Ilichev}},\
  }\bibfield  {title} {\bibinfo {title} {{Quantum and phase diffusion
  crossovers in small Al Josephson junctions}},\ }\href
  {https://doi.org/10.1016/j.chaos.2024.114990} {\bibfield  {journal} {\bibinfo
   {journal} {Chaos, Solitons {\&} Fractals}\ }\textbf {\bibinfo {volume}
  {184}},\ \bibinfo {pages} {114990} (\bibinfo {year} {2024})}\BibitemShut
  {NoStop}%
\bibitem [{\citenamefont {Voss}\ and\ \citenamefont
  {Webb}(1981)}]{Voss1981Jul}%
  \BibitemOpen
  \bibfield  {author} {\bibinfo {author} {\bibfnamefont {R.~F.}\ \bibnamefont
  {Voss}}\ and\ \bibinfo {author} {\bibfnamefont {R.~A.}\ \bibnamefont
  {Webb}},\ }\bibfield  {title} {\bibinfo {title} {{Macroscopic Quantum
  Tunneling in 1-$\mu$m Nb Josephson Junctions}},\ }\href
  {https://doi.org/10.1103/PhysRevLett.47.265} {\bibfield  {journal} {\bibinfo
  {journal} {Phys. Rev. Lett.}\ }\textbf {\bibinfo {volume} {47}},\ \bibinfo
  {pages} {265} (\bibinfo {year} {1981})}\BibitemShut {NoStop}%
\bibitem [{\citenamefont {Martinis}\ and\ \citenamefont
  {Grabert}(1988)}]{Martinis1988Aug}%
  \BibitemOpen
  \bibfield  {author} {\bibinfo {author} {\bibfnamefont {J.~M.}\ \bibnamefont
  {Martinis}}\ and\ \bibinfo {author} {\bibfnamefont {H.}~\bibnamefont
  {Grabert}},\ }\bibfield  {title} {\bibinfo {title} {{Thermal enhancement of
  macroscopic quantum tunneling: Derivation from noise theory}},\ }\href
  {https://doi.org/10.1103/PhysRevB.38.2371} {\bibfield  {journal} {\bibinfo
  {journal} {Phys. Rev. B}\ }\textbf {\bibinfo {volume} {38}},\ \bibinfo
  {pages} {2371} (\bibinfo {year} {1988})}\BibitemShut {NoStop}%
\bibitem [{\citenamefont {Oelsner}\ \emph {et~al.}(2013)\citenamefont
  {Oelsner}, \citenamefont {Revin}, \citenamefont {Il'ichev}, \citenamefont
  {Pankratov}, \citenamefont {Meyer}, \citenamefont {Gr\"{o}nberg},
  \citenamefont {Hassel},\ and\ \citenamefont {Kuzmin}}]{Oelsner2013Sep}%
  \BibitemOpen
  \bibfield  {author} {\bibinfo {author} {\bibfnamefont {G.}~\bibnamefont
  {Oelsner}}, \bibinfo {author} {\bibfnamefont {L.~S.}\ \bibnamefont {Revin}},
  \bibinfo {author} {\bibfnamefont {E.}~\bibnamefont {Il'ichev}}, \bibinfo
  {author} {\bibfnamefont {A.~L.}\ \bibnamefont {Pankratov}}, \bibinfo {author}
  {\bibfnamefont {H.-G.}\ \bibnamefont {Meyer}}, \bibinfo {author}
  {\bibfnamefont {L.}~\bibnamefont {Gr\"{o}nberg}}, \bibinfo {author}
  {\bibfnamefont {J.}~\bibnamefont {Hassel}},\ and\ \bibinfo {author}
  {\bibfnamefont {L.~S.}\ \bibnamefont {Kuzmin}},\ }\bibfield  {title}
  {\bibinfo {title} {{Underdamped Josephson junction as a switching current
  detector}},\ }\bibfield  {journal} {\bibinfo  {journal} {Appl. Phys. Lett.}\
  }\textbf {\bibinfo {volume} {103}},\ \href
  {https://doi.org/10.1063/1.4824308} {10.1063/1.4824308} (\bibinfo {year}
  {2013})\BibitemShut {NoStop}%
\bibitem [{\citenamefont {Blackburn}\ \emph {et~al.}(2016)\citenamefont
  {Blackburn}, \citenamefont {Cirillo},\ and\ \citenamefont
  {Gr{\o}nbech-Jensen}}]{Blackburn2016Feb}%
  \BibitemOpen
  \bibfield  {author} {\bibinfo {author} {\bibfnamefont {J.~A.}\ \bibnamefont
  {Blackburn}}, \bibinfo {author} {\bibfnamefont {M.}~\bibnamefont {Cirillo}},\
  and\ \bibinfo {author} {\bibfnamefont {N.}~\bibnamefont
  {Gr{\o}nbech-Jensen}},\ }\bibfield  {title} {\bibinfo {title} {{A survey of
  classical and quantum interpretations of experiments on Josephson junctions
  at very low temperatures}},\ }\href
  {https://doi.org/10.1016/j.physrep.2015.10.010} {\bibfield  {journal}
  {\bibinfo  {journal} {Phys. Rep.}\ }\textbf {\bibinfo {volume} {611}},\
  \bibinfo {pages} {1} (\bibinfo {year} {2016})}\BibitemShut {NoStop}%
\bibitem [{\citenamefont {Zhou}\ \emph {et~al.}(1990)\citenamefont {Zhou},
  \citenamefont {Moss},\ and\ \citenamefont {Jung}}]{Zhou1990Sep}%
  \BibitemOpen
  \bibfield  {author} {\bibinfo {author} {\bibfnamefont {T.}~\bibnamefont
  {Zhou}}, \bibinfo {author} {\bibfnamefont {F.}~\bibnamefont {Moss}},\ and\
  \bibinfo {author} {\bibfnamefont {P.}~\bibnamefont {Jung}},\ }\bibfield
  {title} {\bibinfo {title} {{Escape-time distributions of a periodically
  modulated bistable system with noise}},\ }\href
  {https://doi.org/10.1103/PhysRevA.42.3161} {\bibfield  {journal} {\bibinfo
  {journal} {Phys. Rev. A}\ }\textbf {\bibinfo {volume} {42}},\ \bibinfo
  {pages} {3161} (\bibinfo {year} {1990})}\BibitemShut {NoStop}%
\bibitem [{\citenamefont {Pankratov}\ and\ \citenamefont
  {Salerno}(2000)}]{Pankratov2000Feb}%
  \BibitemOpen
  \bibfield  {author} {\bibinfo {author} {\bibfnamefont {A.~L.}\ \bibnamefont
  {Pankratov}}\ and\ \bibinfo {author} {\bibfnamefont {M.}~\bibnamefont
  {Salerno}},\ }\bibfield  {title} {\bibinfo {title} {{Adiabatic approximation
  and parametric stochastic resonance in a bistable system with periodically
  driven barrier}},\ }\href {https://doi.org/10.1103/PhysRevE.61.1206}
  {\bibfield  {journal} {\bibinfo  {journal} {Phys. Rev. E}\ }\textbf {\bibinfo
  {volume} {61}},\ \bibinfo {pages} {1206} (\bibinfo {year}
  {2000})}\BibitemShut {NoStop}%
\bibitem [{\citenamefont {Pankratov}\ \emph
  {et~al.}(2022{\natexlab{b}})\citenamefont {Pankratov}, \citenamefont
  {Gordeeva}, \citenamefont {Revin}, \citenamefont {Ladeynov}, \citenamefont
  {Yablokov},\ and\ \citenamefont {Kuzmin}}]{Pankratov2022Jul}%
  \BibitemOpen
  \bibfield  {author} {\bibinfo {author} {\bibfnamefont {A.~L.}\ \bibnamefont
  {Pankratov}}, \bibinfo {author} {\bibfnamefont {A.~V.}\ \bibnamefont
  {Gordeeva}}, \bibinfo {author} {\bibfnamefont {L.~S.}\ \bibnamefont {Revin}},
  \bibinfo {author} {\bibfnamefont {D.~A.}\ \bibnamefont {Ladeynov}}, \bibinfo
  {author} {\bibfnamefont {A.~A.}\ \bibnamefont {Yablokov}},\ and\ \bibinfo
  {author} {\bibfnamefont {L.~S.}\ \bibnamefont {Kuzmin}},\ }\bibfield  {title}
  {\bibinfo {title} {{Approaching microwave photon sensitivity with Al
  Josephson junctions}},\ }\href {https://doi.org/10.3762/bjnano.13.50}
  {\bibfield  {journal} {\bibinfo  {journal} {Beilstein J. Nanotechnol.}\
  }\textbf {\bibinfo {volume} {13}},\ \bibinfo {pages} {582} (\bibinfo {year}
  {2022}{\natexlab{b}})}\BibitemShut {NoStop}%
\bibitem [{\citenamefont {Barends}\ \emph {et~al.}(2011)\citenamefont
  {Barends}, \citenamefont {Wenner}, \citenamefont {Lenander}, \citenamefont
  {Chen}, \citenamefont {Bialczak}, \citenamefont {Kelly}, \citenamefont
  {Lucero}, \citenamefont {O{'}Malley}, \citenamefont {Mariantoni},
  \citenamefont {Sank}, \citenamefont {Wang}, \citenamefont {White},
  \citenamefont {Yin}, \citenamefont {Zhao}, \citenamefont {Cleland},
  \citenamefont {Martinis},\ and\ \citenamefont {Baselmans}}]{Barends2011Sep}%
  \BibitemOpen
  \bibfield  {author} {\bibinfo {author} {\bibfnamefont {R.}~\bibnamefont
  {Barends}}, \bibinfo {author} {\bibfnamefont {J.}~\bibnamefont {Wenner}},
  \bibinfo {author} {\bibfnamefont {M.}~\bibnamefont {Lenander}}, \bibinfo
  {author} {\bibfnamefont {Y.}~\bibnamefont {Chen}}, \bibinfo {author}
  {\bibfnamefont {R.~C.}\ \bibnamefont {Bialczak}}, \bibinfo {author}
  {\bibfnamefont {J.}~\bibnamefont {Kelly}}, \bibinfo {author} {\bibfnamefont
  {E.}~\bibnamefont {Lucero}}, \bibinfo {author} {\bibfnamefont
  {P.}~\bibnamefont {O{'}Malley}}, \bibinfo {author} {\bibfnamefont
  {M.}~\bibnamefont {Mariantoni}}, \bibinfo {author} {\bibfnamefont
  {D.}~\bibnamefont {Sank}}, \bibinfo {author} {\bibfnamefont {H.}~\bibnamefont
  {Wang}}, \bibinfo {author} {\bibfnamefont {T.~C.}\ \bibnamefont {White}},
  \bibinfo {author} {\bibfnamefont {Y.}~\bibnamefont {Yin}}, \bibinfo {author}
  {\bibfnamefont {J.}~\bibnamefont {Zhao}}, \bibinfo {author} {\bibfnamefont
  {A.~N.}\ \bibnamefont {Cleland}}, \bibinfo {author} {\bibfnamefont {J.~M.}\
  \bibnamefont {Martinis}},\ and\ \bibinfo {author} {\bibfnamefont {J.~J.~A.}\
  \bibnamefont {Baselmans}},\ }\bibfield  {title} {\bibinfo {title}
  {{Minimizing quasiparticle generation from stray infrared light in
  superconducting quantum circuits}},\ }\href
  {https://doi.org/10.1063/1.3638063} {\bibfield  {journal} {\bibinfo
  {journal} {Appl. Phys. Lett.}\ }\textbf {\bibinfo {volume} {99}},\ \bibinfo
  {pages} {113507} (\bibinfo {year} {2011})}\BibitemShut {NoStop}%
\bibitem [{\citenamefont {Kurkijärvi}(1972)}]{Kurkijarvi1972Aug}%
  \BibitemOpen
  \bibfield  {author} {\bibinfo {author} {\bibfnamefont {J.}~\bibnamefont
  {Kurkijärvi}},\ }\bibfield  {title} {\bibinfo {title} {{Intrinsic
  Fluctuations in a Superconducting Ring Closed with a Josephson Junction}},\
  }\href {https://doi.org/https://doi.org/10.1103/PhysRevB.6.832} {\bibfield
  {journal} {\bibinfo  {journal} {Phys. Rev. B}\ }\textbf {\bibinfo {volume}
  {6}},\ \bibinfo {pages} {832} (\bibinfo {year} {1972})}\BibitemShut {NoStop}%
\bibitem [{\citenamefont {Fox}(2006)}]{Fox2006Apr}%
  \BibitemOpen
  \bibfield  {author} {\bibinfo {author} {\bibfnamefont {M.}~\bibnamefont
  {Fox}},\ }\href@noop {} {\emph {\bibinfo {title} {{Quantum Optics: An
  Introduction}}}}\ (\bibinfo  {publisher} {OUP},\ \bibinfo {address} {Oxford,
  England, UK},\ \bibinfo {year} {2006})\BibitemShut {NoStop}%
\bibitem [{\citenamefont {Gol{'}tsman}\ \emph {et~al.}(2001)\citenamefont
  {Gol{'}tsman}, \citenamefont {Okunev}, \citenamefont {Chulkova},
  \citenamefont {Lipatov}, \citenamefont {Semenov}, \citenamefont {Smirnov},
  \citenamefont {Voronov}, \citenamefont {Dzardanov}, \citenamefont
  {Williams},\ and\ \citenamefont {Sobolewski}}]{Goltsman2001Aug}%
  \BibitemOpen
  \bibfield  {author} {\bibinfo {author} {\bibfnamefont {G.~N.}\ \bibnamefont
  {Gol{'}tsman}}, \bibinfo {author} {\bibfnamefont {O.}~\bibnamefont {Okunev}},
  \bibinfo {author} {\bibfnamefont {G.}~\bibnamefont {Chulkova}}, \bibinfo
  {author} {\bibfnamefont {A.}~\bibnamefont {Lipatov}}, \bibinfo {author}
  {\bibfnamefont {A.}~\bibnamefont {Semenov}}, \bibinfo {author} {\bibfnamefont
  {K.}~\bibnamefont {Smirnov}}, \bibinfo {author} {\bibfnamefont
  {B.}~\bibnamefont {Voronov}}, \bibinfo {author} {\bibfnamefont
  {A.}~\bibnamefont {Dzardanov}}, \bibinfo {author} {\bibfnamefont
  {C.}~\bibnamefont {Williams}},\ and\ \bibinfo {author} {\bibfnamefont
  {R.}~\bibnamefont {Sobolewski}},\ }\bibfield  {title} {\bibinfo {title}
  {{Picosecond superconducting single-photon optical detector}},\ }\href
  {https://doi.org/10.1063/1.1388868} {\bibfield  {journal} {\bibinfo
  {journal} {Appl. Phys. Lett.}\ }\textbf {\bibinfo {volume} {79}},\ \bibinfo
  {pages} {705} (\bibinfo {year} {2001})}\BibitemShut {NoStop}%
\bibitem [{\citenamefont {Tien}\ and\ \citenamefont
  {Gordon}(1963)}]{Tien1963Jan}%
  \BibitemOpen
  \bibfield  {author} {\bibinfo {author} {\bibfnamefont {P.~K.}\ \bibnamefont
  {Tien}}\ and\ \bibinfo {author} {\bibfnamefont {J.~P.}\ \bibnamefont
  {Gordon}},\ }\bibfield  {title} {\bibinfo {title} {{Multiphoton Process
  Observed in the Interaction of Microwave Fields with the Tunneling between
  Superconductor Films}},\ }\href {https://doi.org/10.1103/PhysRev.129.647}
  {\bibfield  {journal} {\bibinfo  {journal} {Phys. Rev.}\ }\textbf {\bibinfo
  {volume} {129}},\ \bibinfo {pages} {647} (\bibinfo {year}
  {1963})}\BibitemShut {NoStop}%
\bibitem [{\citenamefont {Tucker}\ and\ \citenamefont
  {Feldman}(1985)}]{Tucker1985Oct}%
  \BibitemOpen
  \bibfield  {author} {\bibinfo {author} {\bibfnamefont {J.~R.}\ \bibnamefont
  {Tucker}}\ and\ \bibinfo {author} {\bibfnamefont {M.~J.}\ \bibnamefont
  {Feldman}},\ }\bibfield  {title} {\bibinfo {title} {{Quantum detection at
  millimeter wavelengths}},\ }\href
  {https://doi.org/10.1103/RevModPhys.57.1055} {\bibfield  {journal} {\bibinfo
  {journal} {Rev. Mod. Phys.}\ }\textbf {\bibinfo {volume} {57}},\ \bibinfo
  {pages} {1055} (\bibinfo {year} {1985})}\BibitemShut {NoStop}%
\end{thebibliography}

%apsrev4-2.bst 2019-01-14 (MD) hand-edited version of apsrev4-1.bst
%Control: key (0)
%Control: author (8) initials jnrlst
%Control: editor formatted (1) identically to author
%Control: production of article title (0) allowed
%Control: page (0) single
%Control: year (1) truncated
%Control: production of eprint (0) enabled
%

\end{document}